\documentclass[10pt,journal,compsoc]{IEEEtran}
\usepackage[latin1]{inputenc}
\usepackage{amsmath}
\usepackage{amssymb}
\usepackage{color}
\usepackage{amsfonts}
\usepackage{amssymb}
\usepackage{makeidx}
\usepackage{listings}
\usepackage[pdftex]{graphicx}
\usepackage{times}
\usepackage{algorithm,verbatim,algorithmic}
\usepackage{multirow}
\usepackage{listings}
\usepackage{framed}
\usepackage{sidecap}
\usepackage[hidelinks]{hyperref}
\usepackage[justification=centering]{caption}
\usepackage[nocompress]{cite}
\usepackage[labelfont=bf]{caption}

\usepackage{booktabs,caption,fixltx2e}
\usepackage[flushleft]{threeparttable}

\usepackage{multicol}

\newcommand{\minus}{\scalebox{0.75}[1.0]{$-$}}

\def\denseitems{
  \itemsep1pt plus1pt minus1pt
  \parsep0pt plus0pt
  \parskip0pt\topsep0pt}

\newcommand{\subB}{\vspace{3pt}\noindent\textbf}

\newcommand{\ignore}[1]{}
\definecolor{Light}{gray}{.85}

\def\denseitems{
  \itemsep1pt plus1pt minus1pt
  \parsep0pt plus0pt
  \parskip0pt\topsep0pt}
\begin{document}

\bstctlcite{IEEEexample:BSTcontrol}

\IEEEoverridecommandlockouts
\title{\Name{}: Concurrency Defect Prediction in Real-World Applications}

%
% You need the command \numberofauthors to handle the "boxing"
% and alignment of the authors under the title, and to add
% a section for authors number 4 through n.
%
% Up to the first three authors are aligned under the title;
% use the \alignauthor commands below to handle those names
% and affiliations. Add names, affiliations, addresses for
% additional authors as the argument to \additionalauthors;
% these will be set for you without further effort on your
% part as the last section in the body of your article BEFORE
% References or any Appendices.
%
% You can go ahead and credit authors number 4+ here;
% their names will appear in a section called
% "Additional Authors" just before the Appendices
% (if there are any) or Bibliography (if there
% aren't)

% Put no more than the first THREE authors in the \author command

%% e-mail address with \email. 
%!TEX encoding = UTF-8 Unicode

%\authorinfo {Anonymous}
%       {Computer Science and Engineering\\
%      University of Nebraska-Lincoln\\ 
%     256 Avery Hall\\
%       Lincoln, NE 68588-0115}
%    {Planet Earth}

\newcommand{\Name}{ConPredictor}

\author{Tingting Yu,~\IEEEmembership{Member,~IEEE,}
Wei Wen,  Xue Han, 
Jane Huffman Hayes~\IEEEmembership{Member,~IEEE}% <-this % stops a space
\IEEEcompsocitemizethanks{\IEEEcompsocthanksitem All authors are
with the Department
of Computer Science, University of Kentucky, Lexington, Kentucky, USA, 40506.\protect\\
E-mail: tyu@cs.uky.edu,
wei.wen0@uky.edu,
xha225@g.uky.edu,
hayes@cs.uky.edu}
\thanks{Manuscript received June 27, 2017; revised November 16, 2017.}
}

\IEEEtitleabstractindextext{%
\begin{abstract}
Concurrent programs are difficult to test due to their inherent non-determinism.
To address this problem, testing often
requires  the  exploration  of  thread  schedules  of  a  program;  this  can
be  time-consuming  when applied to real-world  programs.  
Software defect prediction has been used to help developers find faults
and prioritize their testing efforts.   
Prior studies have used machine learning to build such predicting models based on designed features that encode
the characteristics of programs.  
However, research has focused
on  sequential  programs; to  date,  no  work  has  considered  defect
prediction for concurrent programs, with program characteristics distinguished from sequential programs. In this paper, we present \Name{}, 
an approach to predict defects specific to 
concurrent programs by combining both static and dynamic program metrics. Specifically, 
we propose a set of novel static code metrics based on the unique properties of concurrent programs.
We also leverage additional guidance from dynamic metrics constructed 
based on mutation analysis. 
Our evaluation on four large open source projects shows that \Name{} improved 
both within-project defect prediction and 
cross-project defect prediction  compared to traditional features.
\end{abstract}

}

\maketitle

\IEEEdisplaynontitleabstractindextext

\IEEEpeerreviewmaketitle

\IEEEraisesectionheading{\section{Introduction}}
\label{sec:intro}

\IEEEPARstart{S}{oftware} quality assurance is an expensive activity:  it requires time and resources to be performed properly, and it delays a product's delivery to market.  
%because the time-to-market requirements of software delivery must be met. 
This high-cost issue is more challenging in 
many of today's concurrent software systems due to their complicated
behaviors.  For example, assuring the quality of concurrent programs is difficult
 primarily because testing faces this challenge:  concurrency faults are sensitive to execution interleavings
that are imposed by various concurrency constructs (e.g., synchronization operations). 
Testing usually requires exploring as many interleavings as possible to amplify 
the chance of exposing faults. 
Recent work~\cite{Deng13} reports that testing concurrent programs 
can introduce a 10x-100x slowdown for each test run.  
Such overhead increases as test suite size increases.
Therefore, it is desirable to determine which code regions are more
likely to contain concurrency faults as this can guide developers
to focus the testing efforts on the identified code, thus reducing the time and resources required for testing and leading to reduced quality assurance costs. 

For this reason, defect prediction has been an active research area in 
software engineering~\cite{khoshgoftaar1995application,li2006experiences, Lessmann08}.
Defect prediction techniques build models from software data and use the models to predict 
whether new instances of code regions, e.g., files, changes, and methods, contain defects. 
These techniques  first design features or combinations of features and then use
machine learning algorithms to build prediction models.
Based on the prediction results, developers can allocate limited testing 
efforts more effectively to focus on the defect-prone modules.
In particular, source code metrics have been used
widely and shown effectiveness in prediction. 
There has been much research on software defect prediction by combining
static code metrics to identify defect-prone source code artifacts~\cite{Zhang-2007}. 
A variety of statistical and machine learning techniques have been used to build defect prediction 
models~\cite{WORDS-2005}.

However, all existing defect prediction research has focused on sequential
software.  To date, no work has considered concurrent software systems for which
adapting existing prediction models may not be effective. 
Unlike defect prediction for sequential programs, that often relies
on a set of well-defined and traditional code metrics (e.g., lines of code,
cyclomatic complexity), predicting concurrency defects in concurrent programs 
must consider the unique concurrency properties in its fault models:  
threads, shared variable accesses between threads, and synchronization operations. 

Another challenge is that most existing research has focused 
on designing features from static code analysis. 
However, the performance of defect prediction  can also rely on the
quality of the test suite~\cite{Bowes2016}. This is because
the performance of a defect prediction model should 
be measured as its ability to predict faults that ultimately lead
to test failures. Although recent work~\cite{Bowes2016} has proposed
using mutation analysis to guide defect detection, it focuses on
sequential programs.  None of the existing research has considered
using either static or dynamic metrics to predict defects
for concurrent programs. 

In this paper, we propose \Name{}, a defect prediction framework
for predicting functions that are likely to contain concurrency defects in
real-world applications. 
Specifically, we  propose six novel code metrics 
specific to concurrent programs  by taking unique features related to 
concurrency properties into account. 
We adapt the concurrency control flow graph (CCFG) to generate code
metrics involving:  (i) concurrent cyclomatic complexity, (ii)
number of shared variables, (iii) number of conditional basic blocks that contain 
concurrency constructs, (iv) number of communication edges in CCFG, 
(v) number of synchronization operations,
and (vi) access distance between shared variables in a local thread. 
We then define 18 mutation metrics computed by applying a variety of mutation 
operators specific to concurrent programs.
These metrics include six static metrics (e.g., the number of times
a mutation operator is applied) and 12 dynamic metrics from dynamic mutation analysis. 
The \Name{} prediction model is built upon all 24 static and dynamic concurrency metrics. 
Then, we empirically compare the performance of \Name{} to those
of prediction models built using traditional metrics that have been 
widely used in previous fault prediction work~\cite{Lessmann08}.
We also investigate whether the combined use of mutation metrics and source code metrics improves 
the accuracy of the resulting prediction model. 
Moreover, we examine the extent to which different machine learning techniques
benefit from the metrics in \Name{}.  We also determine the best combination
of metrics for predicting testability.

To evaluate our approach, we apply it to four large real-word systems. 
Our primary finding is that \Name{} can significantly improve the prediction performance, with large effect sizes, when comparing to the 
traditional metrics for sequential programs. The dynamic concurrency 
metrics are even more effective than the static concurrency metrics. 
Our paper makes the following contributions:

\begin{enumerate}\denseitems
\item 
The first approach to effectively predict concurrency faults,
\item 
A set of novel source code metrics specific to concurrent programs, 
\item
The introduction of dynamic concurrency metrics for fault prediction, and
\item
An empirical study showing the effectiveness of our approach. 
\end{enumerate}

This article draws on our previously published 
conference paper~\cite{yu2016predicting}, 
in which we propose a set of novel static code metrics to 
predict testability of concurrent programs. 
The testability is calculated based on mutation analysis. 
We have extended this work substantially by  
introducing a new static
metric and a set of dynamic concurrency metrics.
We then use both static and dynamic metrics to predict 
concurrency faults. All concurrency faults considered
in this work are real. 
We have performed an empirical study that is
different from our previous work in terms
of research questions, study design,
results, and analysis. 
\footnote{Note to
reviewers: our cover letter provides more details 
on the extensions and revisions that differentiate
this work from the original conference papers.}

The remainder of the paper is organized as follows.  Section \ref{background} presents background and definitions.  
Our approach and concurrency code metrics are introduced in Section \ref{CPM}.  Section \ref{study} discusses the 
empirical study.  Results are presented in Section \ref{results} followed by discussion in Section \ref{discussion}.  
Prior work is presented in Section \ref{related} and Section \ref{conclusions}
concludes.

\section{Background and Definitions} \label{background}

In this section, we provide background information on 
defection prediction and mutation analysis. 
Related work is discussed further in Section \ref{related}.

\subsection{Software Defect Prediction}

\begin{figure}[t]
%\hspace{5pt}
\includegraphics[clip=true,width=3.5in]{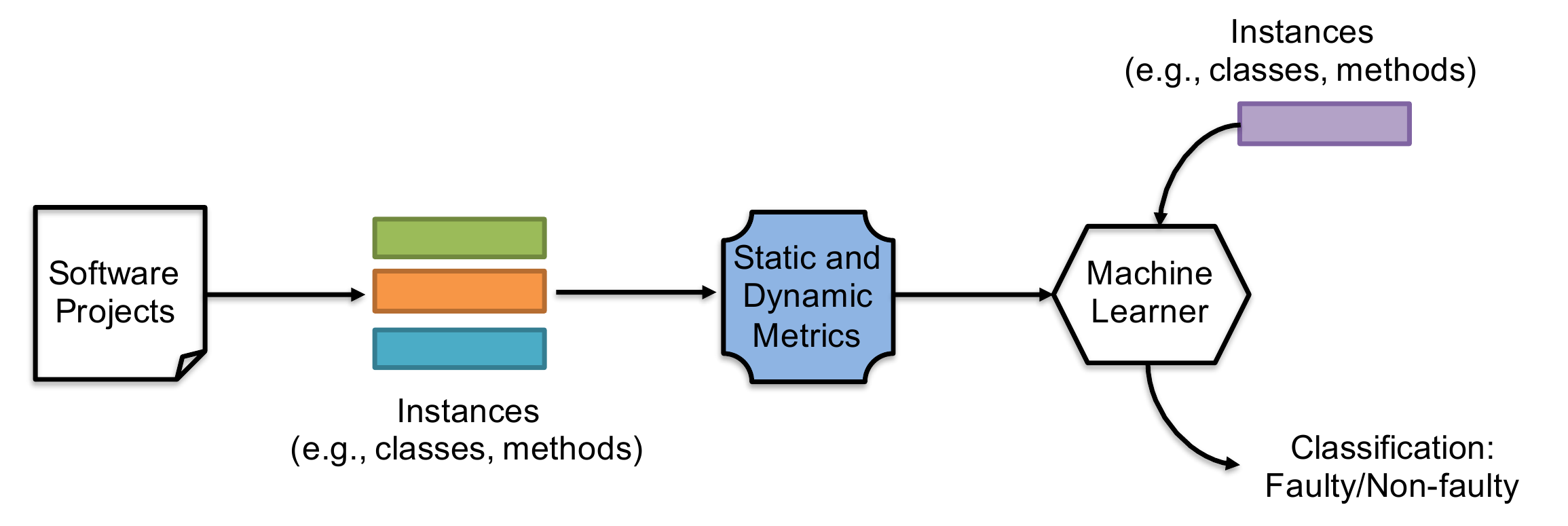}
%\vspace*{-12pt}
\centering{
\caption{\textbf{\small Process of Predicting Software Faults}}
\label{fig:learning}
}
%\vspace*{-12pt}
\end{figure}

A software defect prediction model generally
exploits historical data 
%about software modules 
to 
classify software modules as either faulty or non-faulty.
A prediction model infers a  single aspect of the data (i.e., dependent variable) from a
combination of other aspects of the data (e.g., independent variables).
In the software fault prediction context, the dependent variable is 
the label indicating whether a software module contains a fault or not 
while the independent variables can be related to different aspects of 
the software such as source code metrics.
Figure~\ref{fig:learning} illustrates the process of defect prediction. 

The performance of the fault prediction model depends both on the modeling
technique and the independent variables (i.e., metrics) used. 
Classification techniques (e.g.,
Decision Trees, Logistic Regression, and Naive Bayes)
have been widely used to build fault predication
models~\cite{Lee11,Menzies10,harrold01mar}.
However, according to recent systematic literature
reviews~\cite{hall2012systematic,radjenovic2013software}, 
the choice of a modeling technique seems to have less impact on the 
classification accuracy of a model than the choice of a metrics set. 

Feature selection has been used to select a set of most relevant independent
variables contained in the original dataset to eliminate variables that do not
contribute to the performance of prediction, and can thus improve
learning efficiency and increase prediction accuracy~\cite{harman2014angels}. 
Feature selection is usually performed by leveraging a machine learning algorithm that 
can evaluate the usefulness of the feature set (i.e.,
wrappers~\cite{kohavi1997wrappers}).  
This can also be done by ranking methods
(i.e., filters) that evaluate the features according to heuristics 
based on general characteristics of the data. 

The performance of a classification model is typically evaluated based on the confusion matrix. 
The matrix contains four instances: True Positive (TP) -- 
faulty components correctly classified as faulty; False Negative (FN) -- 
faulty components incorrectly classified as non-faulty; False Positive (FP) -- 
non-faulty components incorrectly classified as faulty; and True Negative (TN) -- 
non-faulty components correctly classified as non-faulty. The confusion matrix 
values are used to calculate a set of evaluation measures, including
precision (measuring the proportion of the components classified 
as faulty which are actually faulty), recall (measuring the proportion of faulty components 
classified as faulty), and F-Measure (which is the harmonic mean of precision and recall).

\subsection{Mutation Analysis}
\label{ma}

Mutation testing is an approach for evaluating 
test suites and testing techniques using a large number of systematically seeded program changes, 
allowing for statistical analysis of results~\cite{Ammann08,Jia11}. 
Mutation testing typically involves three stages. (1) Mutant generation -- 
in this stage, a predefined set of mutation operators are used to generate
mutants from program source code or bytecode.
A mutation operator is a rule that is applied to a program to create 
mutants, such as arithmetic operator replacement (AOR) ~\cite{Wong01}. 
(2) Mutant execution -- 
in this stage, the goal is execution of test cases 
against both the original program and the mutants. (3) Result analysis -- 
in this stage, 
the goal is to check the mutation score obtained by the test suite, where mutation score is defined as the
ratio of the number of killed mutants to the number of all (non-equivalent) 
generated mutants.

Mutants that contain a single fault are called first-order mutants. First-order
mutants have been widely used for mutation analysis of sequential programs. 
Table~\ref{tab:seq-operators} lists six commonly used mutation operators for
sequential programs~\cite{agrawal1989design}, including operator
indexes, operator names, and descriptions. 
These operators are used in our study.

There has been some work to generate first-order mutants for concurrent
programs~\cite{Ghosh02,Bradbury06}. 
For example, Ghosh generates concurrency-related mutants by removing
single synchronization keywords~\cite{Ghosh02}. 
Bradbury et al. proposed a  set of first-order mutation operators for Java~\cite{Bradbury06}. 
However, more recent work has shown that first-order mutants are not sufficient
to simulate subtle concurrency faults due
to the complexity of thread synchronizations~\cite{Gligoric13, Kaiser11,Kusano13}.
%In fact, the majority of the mutants are generated by a small subset of 
%the mutation operators, generally those that are synchronization-centric:  directly related to the synchronization of different processes or threads. 
Therefore, some research has investigated higher-order mutants~\cite{Harman11, Jia09}
for concurrent mutation operators~\cite{Kaiser11} by inserting
two or more faults. Higher-order mutants subsume first-order mutants, as killing the former
is a sufficient but not necessary condition for killing the latter. 

\begin{table}[t]
\centering
\footnotesize
\caption{\label{tab:seq-operators} \textbf{\small List of Sequential Mutation
Operators}}
%\vspace*{-6pt}
%\setlength{\tabcolsep}{1pt}
\begin{tabular}{|l|l|l|} 
%\hline
%\multicolumn{3}{|c|}{Sequential Mutation Operators} \\
\hline
index &{\em Operator} & {\em Description} \\
\hline
%1 & SMTC & n-trip continue\\
%\hline
1 & ssdl & statement deletion\\
\hline
%3 & STRP & trap on statement execution\\
%\hline
2 & swdd & while replacement by do-while\\
\hline
3 & oasn & arithmetic operator by shift operator\\
\hline
4 & oeba & plain assignment by bitwise assignment\\
\hline
%7 & OLBN & logical operator by bitwise operator\\
%\hline
%8 & OLLN & logical operator mutation\\
%\hline
5 & olng & logical negation\\
\hline
6 & orrn & relational operator mutation\\
\hline
%11 & VTWD & twiddle mutations\\
%\hline
%12 & VDTR & domain traps\\
%\hline
%13 & Cccr & constant for constant replacement\\
%\hline
%14 & Ccsr & constant for scalar replacement\\
%\hline
%15 & CRCR & required constant replacement\\
%\hline
\end{tabular}
\normalsize
%\vspace*{-12pt}
\end{table}

\begin{table}[t]
\centering
\footnotesize
\caption{\label{tab:con-operators} \textbf{\small List of Concurrency Mutation
Operators}}
%\vspace*{-6pt}
%\setlength{\tabcolsep}{1pt}
\begin{tabular}{|l|l|l|} 
%\hline
%\multicolumn{3}{|c|}{Sequential Mutation Operators} \\
\hline
index &{\em Operator} & {\em Description} \\
\hline
1 & rmlock & Remove call to lock/unlock\\
\hline
%2 & msem & Modify permit count in semaphore\\
%\hline
%2 & mwait & Modify parameter/time in cond\_timedwait\\
%\hline
2 & rmwait & Remove call to cond\_wait/cond\_timedwait\\
\hline
%5 & swptw & Swap cond\_wait with cond\_timedwait\\
%\hline
3 & rmsig & Remove call to cond\_signal/cond\_broadcast\\
\hline
%4 & swptw & Swap cond\_signal with cond\_broadcast\\
%\hline
4 & rmjoinyld & Remove call to join/yield \\
\hline
%9 & repjn & Replace join with sleep\\
%\hline
%10 & rmvol & Remove volatile keyword\\
%\hline
%11 & swplck & Swap lock-unlock pairs\\
%\hline
5 & shfecs & Shift critical section\\
\hline
%13 & shkecs & Shrink critical section \\
%\hline
%14 & epdecs & Expand critical section\\
%\hline
6 & spltecs & Split critical section\\
\hline
%16 & rmbarrier & Remove call to barrier\_wait\\
%\hline
%17 & mbarrier & Modify parameter/time in barrier\_wait\\
%\hline
\end{tabular}
\normalsize
%\vspace*{-12pt}
\end{table}

To generate higher order mutants for concurrent programs, 
Kaiser et al.~\cite{Kaiser11}  propose a set of mutation operators
for multi-threaded Java programs based on concurrency 
bug patterns that include subtle concurrency faults (e.g., data races).
Kusano et al.~\cite{Kusano13} implemented {\em CCmutator} 
based on the Clang/LLVM compiler framework to inject
concurrency faults for multithreaded C/C++  applications.
Their work considers both first-order and higher-order mutants. 
In this work, we consider 
various  concurrency-related mutation operators 
from {\em CCmutator}.
Table~\ref{tab:con-operators} summarizes the mutation operators used
in this paper.
%, including operator names, descriptions, and their possible
%consequences~\cite{Lu08}.  

These operators include
mutex locks, condition variables, atomic objects, semaphores, 
thread creation, and thread join. For example, removing a lock-unlock
pair can create potential data races and atomicity violations, 
removing a conditional wait can create order violations,
shifting and splitting critical sections can introduce potential data races
and order violations as some variables are no longer synchronized. 
Replacing a call to join with a call to sleep can cause nondeterministic 
behavior. 
%If the sleep time is sufficiently long, the program may appear to be correct. 
%Otherwise, the program may crash due to the improper join. 

We use the term location to indicate a place in a program where a fault can 
occur. Although the techniques we propose can be used at different granularities (e.g., statements),
this paper concentrates on locations that correspond to single program instructions.

\section{Concurrency-related Code Metrics}
\label{CPM}

Figure~\ref{fig:learning} shows the process of fault prediction in this work.
First, we define instances as units of programs, 
these can be files, classes, or functions.
The instances that we consider are at the function level.
We label an instance as faulty if it has any concurrency faults,
or non-faulty otherwise. 
The next step is to compute static and dynamic metrics.
The static metrics consist of static code metrics and static 
mutation metrics, and the dynamic metrics consist of coverage
metrics and dynamic mutation metrics. 
To obtain static code metrics, we perform static analysis 
on the concurrency control flow graph (CCFG).
% to extract static code metrics. 
We obtain the static mutation metrics
by recording the number of times each mutation operator is applied.
We then execute test cases on
the instances to compute metrics based on code coverage.
Next, we apply concurrency mutation operators
to the instances and execute test cases to collect dynamic mutation metrics.
%Each mutation operator associates with a mutation metric.
%In the meantime, we also compute test suite metrics based on code coverage.
Finally, we train prediction models using machine 
learning algorithms implemented in Weka~\cite{Hall09}. 
The trained prediction models classify instances as faulty or 
non-faulty.

In the following sections, we describe the approach
to computing the static and dynamic metrics 
for concurrency fault prediction.

\subsection{Concurrency Control Flow Graph}

\begin{figure*}[t] \label{ccfg}
%\hspace{5pt}
\includegraphics[clip=true, scale=0.6]{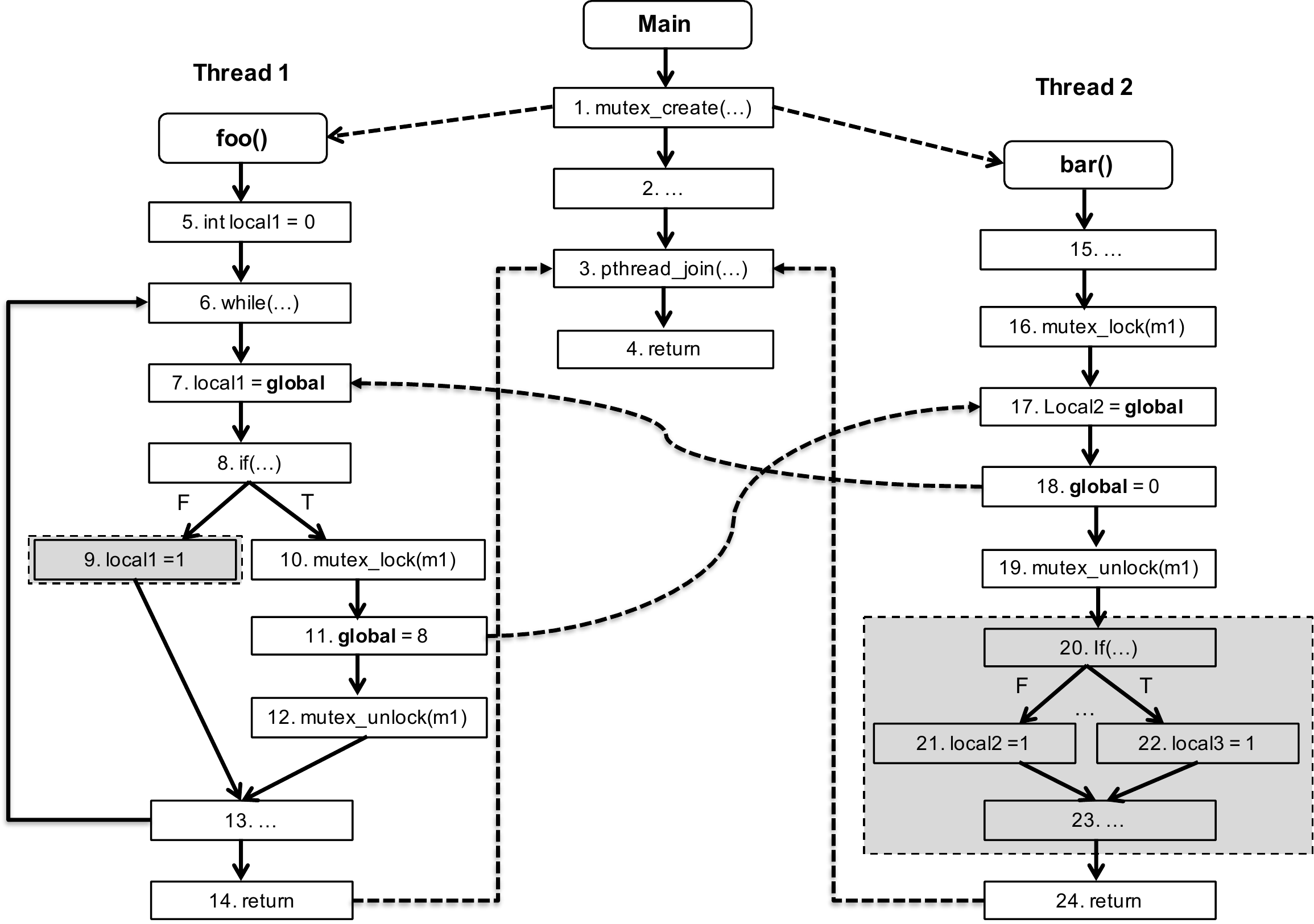}
%\vspace*{-12pt}
\centering{
\caption{\textbf{Concurrency Control Flow Graph}}
\label{fig:ccfg}
}
%\vspace*{-12pt}
\end{figure*}

A concurrent program $P$ consists of threads that communicate with each other  
through shared variables and synchronization operations. 
Given the program source code, we can construct a concurrent 
control flow graph (CCFG) for a procedure $p \in P$
based on a flow and context-sensitive pointer analysis,
where $p$ can be accessed by multiple threads. The idea of building
CCFGs is not new and there has been research on using CCFGs to achieve
different objectives~\cite{Ganai10, Kahlon09}. For example, Kahlon 	
et al.~\cite{Kahlon09}  build a context-sensitive CCFG to perform staged data race detection. 
Our CCFG is similar to those used in existing work but is 
implemented to satisfy our goal of predicting concurrency faults. 
First, $p$ is constructed into a control flow graph (CFG), denoted
as ($N(p)$, $E(p)$). A node $N(p)$ is an instruction $I$ 
and an edge $I_i \rightarrow I_j \in E(p)$ describes the control flow
and data flow between nodes in this CFG.  
In the CCFG, we add additional edges to represent communications between procedures
potentially running on two different threads. Figure~\ref{fig:ccfg} illustrates 
an example CCFG, where the solid lines reflect the local edges and the dotted 
lines reflect the cross-threads'
edges, including {\tt fork}, {\tt join}, and {\tt communication} edges.  
The {\tt Main} function creates two 
threads, on which functions {\tt foo} and {\tt bar} are running, respectively. 
The variables marked as bold are shared between threads.
For readability purposes, we use statements rather than instructions to represent each node.  

Specifically, a {\tt fork} edge is added from the 
program location where {\tt thread\_create} instruction is called to the
entry node of the procedure to be executed. 
In Figure~\ref{fig:ccfg}, edges $<1, foo>$ and $<1, bar>$ form two fork edges. 
If the thread on which the procedure 
is to be executed is specified as in the thread pool model, the procedure is duplicated
on the other thread.  A {\tt join} edge is added from the return of a procedure
that is executed by the fork to the node representing {\tt thread\_join} instruction. 
In Figure~\ref{fig:ccfg}, edges $<14, 3>$ and $<24, 3>$ are considered to be join edges. 
A {\tt communicate} edge is added from a write of one shared variable (SV)  on one thread
to the read of the same SV on the other thread. For example, Figure~\ref{fig:ccfg}
contains two communication edges involving two shared variable pairs
$<11, 17>$ and $<18, 7>$.

\subsection{Static Metrics}

We introduce five code metrics 
and ten mutation metrics specific to concurrent programs. 
%We implemented CCFG on top of the LLVM framework \cite{llvm}, so the code metrics are generated at the instruction level.

\subsubsection{Static Code Metrics}

Static code metrics are generated from the CCFG. 

\subB{Synchronization point count (SPC).}
We define the {\em synchronization point count}, $SPC(f)$,
as the number of nodes involving synchronization operations ($SO$s) in a function $f$. 
The use of the SPC metric is based on the intuition that
the number of $SO$s contributes to 
the complexity of concurrent programs. As the 
number of $SO$s increase, 
the program is more likely to contain more faults
related to synchronization usage, such as deadlock and
atomicity violations. The SPC metric for a function is defined as:  

\begin{center}
$SPC(f) = num\_of\_syncs(f)$ 
\end{center}

\noindent
Here, we consider mutex, semaphore, conditional variables,
and barriers as synchronization operations. In the Figure~\ref{fig:ccfg} example, $SPC(main)$ = 2,
$SPC(foo)$ = 2, and $SPC(bar)$ = 2, because
each function contains two synchronization operations (e.g., {\tt mutex\_lock}).

\subB{Shared variable count (SVC).}
In this metric, we count the nodes in the CCFG involving 
shared variable (SV) read/write in the procedure $p$. 
The shared variables can affect data communication
between threads. 
The increasing complexity of SV usage is likely to 
cause incorrect data state to propagate across threads. 
As such, we define SVC for $p$ as:

\begin{center}
$SVC(f) = num\_of\_SVs(f)$
\end{center}

\noindent
In Figure~\ref{fig:ccfg},  $SVC(foo)$ = 2 and $SVC(bar)$ = 2.
The variables marked with bold are SVs. Note that we do not
count SVs passed as lock objects.
$SVC(main)$ = 0, because there are no global variables
in the {\tt main} function.

\subB{Conditional synchronization count (CSC).}
We define {\em conditional synchronization count}, $CSC(f)$,
as the number of conditional basic blocks (e.g., branches)
within function $f$ that contain at least one shared variable
or synchronization operation. The CSC metric takes into 
account the local control flow of a procedure in one thread that can potentially
complicate the communication with procedures running on other threads. 
The intuition is that the conditional block containing concurrency
elements increases the complexity of a function in terms
of multithreading communication, affecting the sensitivity
of inputs that reach such synchronization points. 
The CSC metric for $f$ is defined as: 

\begin{center}
$CSC(f) =num\_of\_cond\_syncs(f)$ 
\end{center}

\noindent
As Figure~\ref{fig:ccfg} shows, $CSC(foo)$ = 2, because
there are two conditional blocks in {\tt foo} that contain SVs 
(i.e., {\tt while}, {\tt if}). The shaded areas are irrelevant basic 
blocks that CSC does not count. Note that $CSC(bar)$ = 0 and $CSC(main)$ = 0,
because the two functions contain no conditional basic blocks containing
SVs or synchronization points 
(the two shaded conditional basic blocks in {\tt bar} are irrelevant basic blocks).

\subB{Communication edge count (CEC).}
The {\em communication edge count} metric, denoted
by $CEC(f)$, counts the number of communication
edges involved across all shared variables in a
function $f$. Since communication edges indicate how
different threads can interleave, CEC reflects
the complexity of interleaving space. 
In this case, the CEC value may be proportional to 
the likelihood of exposing concurrency faults.

\begin{center}
$CEC(f) =num\_of\_comm\_edges(f)$ 
\end{center}

\noindent

In Figure~\ref{fig:ccfg}, $CEC(main)$ = 4, $CEC(foo)$ = 4,
and $CEC(bar)$ = 4. Each of the four functions contains
four communication edges.

\subB{Concurrency cyclomatic complexity (CCC).}
We extend the traditional McCabe's cyclomatic complexity \cite{McCabe}
to measure complexity of concurrent programs, 
denoted as  {\em concurrency cyclomatic complexity} (CCC). 
To compute CCC, we first prune the CCFG to transform it into CCFG', 
which involves two steps:
1) remove each basic block $b$ that is irrelevant to 
concurrency properties (i.e., SVs and synchronizations) computed
by the $CSC$ metric, as well as remove their incoming and outgoing edges; and
2) add an auxiliary edge from each predecessor of $b$
to each successor of  $b$. Thus, the $CCC(p)$ of a 
function is defined with reference to its CCFG':

\begin{center}
$CCC(f) = E' - N' + 2$
\end{center}

\noindent
Here, $E'$ is the number of edges 
and $N'$ is the number of nodes in the $f$ of CCFG,
where $E'$ includes both ingoing and outgoing edges for $f$.
In Figure~\ref{fig:ccfg}, 
the shaded nodes are irrelevant and thus removed.
Auxiliary edges are added from node 8 to node 13, and
from node 19 to node 24. Thus, $CCC(main)$ = 8 - 5 + 2 = 5, 
$CCC(foo)$ = 15 - 10 + 2 = 7, and $CCC(bar)$ = 10 - 7 + 2 = 5. 
Note that the sequential Cyclomatic Complexity 
of the three functions are 5 - 4 + 1 = 2, 12 - 11 + 2 = 3, and
11 - 11 + 2 = 2. 

\subB{Shared variable access distance (SVD).}
The  distance between two shared variable ($SV$) accesses 
in one thread is also an important factor for concurrency fault
exposure~\cite{Lu07}. For example, if {\tt x} is written at $l_1$ and later 
read at $l_2$ by the same thread $T1$, and there exists a different thread $T2$
that updates the value of {\tt x}, the distance between $l_1$
and $l_2$ can impact the chances of $T2$ interleaving between them.
We consider access distance as the instruction 
gap between two $SV$ accesses (reads or writes)
in the same procedure $p$.  To compute $SVD(f)$, we first
identify all SV pairs ($SVP$) in $p$. For each $SV$ pair $<sv1, sv2>$, we calculate 
all instruction gaps by traversing all path segments\footnote{A path segment is a path slice for which every node is visited at most once.}
between $sv1$ and $sv2$. Note that one $SVP$ can associate with multiple distance values
due to possible control flow edges. To consider all path segements, 
we calculate the mean distance over all path segments.  
Specifically, we average all distance values if they are normally distributed, 
otherwise we use their trimmed mean. Thus, $SVD(f)$ is defined as:

\begin{center}
$SVD(f)$ = $\overline{\sum\limits_{i=1}^{N}{\sum\limits_{i=j}^{M} Dis(SVP_{i, j})}}$
\end{center}

\noindent

Here, $N$ is the number of shared variable pairs and $M$ is the number of
path segments for an $SVP$. 
In the function {\tt foo} of Figure~\ref{fig:ccfg}, there are
two path segments: (7,8,10,11) and (11,12, 13, 6,7) for $SVP$s $<$7, 11$>$ and 
$<$11, 7$>$, respectively. Suppose each node counts for 2 instructions, then
$SVD(foo)$ = (8 + 10) / 2 = 9.

\subsection{Mutation Metrics}
\label{dm}

Most previous fault prediction studies have used 
static code metrics for this prediction task. Dynamic information resulting from test executions
can also be leveraged for guiding fault prediction.
In this paper, we use mutation testing to produce
both static and dynamic metrics.
% that can guide defect prediction. 
We choose mutation testing for two reasons. 
First, the quality of the test suite is a considerable 
factor for finding faults: 
mutation testing has been used to assess the quality of
a test suite~\cite{demillo1978hints,king1991fortran}.
Second, empirical study~\cite{Andrews:ICSE05,JustJIEHF2014} 
has shown that test cases that are good at killing 
mutants are good at detecting real faults.
Therefore, it has been validated that mutants  
can be a legitimate substitute for real faults in testing. 

We consider three metric suites: one static metric
suite and two dynamic metric suites.

\subsubsection{Static Mutation Metrics}
The concurrency static mutation metric suite
consists of a list of 
mutation metrics  that simply record the number of 
times a particular mutation operator can be applied.
They are static metrics because 
%they are independent of any test suite.
the metric data can be obtained statically without
executing the code.  Therefore, these
metrics are independent of any test suites. 
Each mutation operator is associated with a metric. 
For instance, the six concurrency mutation metrics
listed in Table~\ref{tab:con-operators} yield six static mutation metrics. 

\subsubsection{Dynamic Mutation Metrics}
The dynamic mutation metrics take into account the test suite as well 
as the potential faults that can be seeded.
For each type of concurrency mutation operator defined in Section~\ref{ma}, 
we compute two corresponding
dynamic metrics:  
the percentage of mutants 
(with respect to a mutation operator) that are executed (MuDuE),
and the percentage of mutants 
(with respect to a mutation operator) that are killed (MuDuK).
This results in 12 (2 $\times$ 6) dynamic metrics,
each of which contributes to an independent variable 
of the prediction model. 
Each metric can also assess the quality of the test suite
regarding
whether or not the test suite can exercise 
a particular simulated fault (covered mutant), and whether 
the test suite can detect this simulated fault (killed mutant).

\subsection{Implementation}
Our metrics are implemented using the popular Clang/LLVM compiler platform~\cite{llvm} using
the LLVM opt pass~\cite{llvmpass} to collect program information. Clang's CFG provides
a directed graph for each function, where the nodes are the basic blocks and the directed edges represent how the control flows. As noted above, our CCFG extends
the basic CFG by adding edges describing inter-thread communication. 
We apply shared variable analysis to identify variables
shared by two threads, such as heap objects and data objects that are passed to a function 
(e.g., thread starter function) called by another thread.
Since our metrics implemented by LLVM's intermediate representation (IR) are based on single static 
assignment (SSA) form, we can potentially  leverage compiler front-ends to handle other languages. 
We leverage CCMutator~\cite{Kusano13}, a concurrency mutation tool, to generate mutants of concurrent
programs.

\section{Empirical Study}
\label{study}

\begin{table*}[t]
\centering
\footnotesize
\caption{\label{tab:metrics} \textbf{\small List of Static Code Metrics}}
%\vspace*{-6pt}
%\setlength{\tabcolsep}{1pt}
\begin{tabular}{|l|l|l|l|} \hline
Predictor & Metric Suite & & \\
\hline
\multirow{24}{*}{ConPredictor}& & {\em Concurrency Code Metrics (CCM)}  & {\em
Description}\\
\cline{3-4}
& \multirow{12}{*}{CStaMc} & ConcurrencyComplexity(CCC) & Concurrent
program complexity\\
\cline{3-4}
& &CountSharedVaraible (CSV) & \# of shared variables\\
\cline{3-4}
& &CountConditionalBasicBlock (CBB) & \# of conditional basic blocks\\
\cline{3-4}
& &CountSynchronizations(CSO) & \# of synchronization operations\\
\cline{3-4}
& &CountComEdges(CCE) & \# of communication edges\\
\cline{3-4}
& &CountDistance(CCD) & average distance between shared variables\\
\cline{3-4}
\cline{3-4}
& &{\em Con. Stat. Mut. Met. (CSMM)} & {\em Description}\\
\cline{3-4}
& &MuS$_{rmlock}$ & \# of generated mutants on $rmlock$\\
\cline{3-4}
& &MuS$_{rmwait}$ & \# of generated mutants on $rmwait$\\
\cline{3-4}
& &MuS$_{rmsig}$ & \# of generated mutants on $rmsig$\\
\cline{3-4}
& &MuS$_{rmjoinyld}$ & \# of generated mutants on $rmjoinyld$\\
\cline{3-4}
& &MuS$_{shfecs}$ & \# of generated mutants on $shfecs$\\
\cline{3-4}
& &MuS$_{spltecs}$ & \# of generated mutants on $spltecs$\\
\cline{3-4}
\cline{2-4}
& &{\em Con. Dyn. Mut. Met. (CDMM)} & {\em Description}\\
\cline{3-4}
&\multirow{12}{*}{CDyMc} &MuDuE$_{rmlock}$ & \% of mutants executed on
$rmlock$\\
\cline{3-4}
& &MuDuE$_{rmwait}$ & \% of mutants executed on $rmwait$\\
\cline{3-4}
& &MuDuE$_{rmsig}$ & \% of mutants executed on $rmsig$\\
\cline{3-4}
& &MuDuE$_{rmjoinyld}$ & \% of mutants executed on $rmjoinyld$\\
\cline{3-4}
& &MuDuE$_{shfecs}$ & \% of mutants executed on $shfecs$\\
\cline{3-4}
& &MuDuE$_{spltecs}$ & \% of mutants executed on $spltecs$\\
\cline{3-4}
& &MuDuK$_{rmlock}$ & \% of mutants killed on $rmlock$\\
\cline{3-4}
& &MuDuK$_{rmwait}$ & \% of mutants killed on $rmwait$\\
\cline{3-4}
& &MuDuK$_{rmsig}$ & \% of mutants killed on $rmsig$\\
\cline{3-4}
& &MuDuK$_{rmjoinyld}$ & \% of mutants killed on $rmjoinyld$\\
\cline{3-4}
& &MuDuK$_{shfecs}$ & \% of mutants killed on $shfecs$\\
\cline{3-4}
& &MuDuK$_{spltecs}$ & \% of mutants killed on $spltecs$\\
\hline
\hline
& &{\em Sequential Code Metrics (SCM)}
& {\em Description}\\
\cline{3-4}
%\multirow{24}{*}{SPM}& &CountInstruction (CI)~\cite{nagappan2006mining} & \# of instructions\\
%\cline{3-4}
%& \multirow{12}{*}{SStaMc} &CountBasicBlock (CB)~\cite{nagappan2006mining} & \#
%of basic blocks\\
%\cline{3-4}
%& &CountParameter(CPA)~\cite{nagappan2006mining} & \# of parameters for a
%function\\
%\cline{3-4}
%& &CyclomaticComplexity(CC)~\cite{mccabe1976complexity} & Mcabe's cyclomatic
%complexity\\
%\cline{3-4}
%& &CountFunctionCall(CM)~\cite{Lee11} & \# of function calls\\
%\cline{3-4}
%& &CountVariables(CV)~\cite{nagappan2006mining} & \# of variables\\
\multirow{27}{*}{SPM}& & CountFunctionIn (CN) & \# of functions that call a given function\\
\cline{3-4}
& \multirow{15}{*}{SStaMc} & CountFunctionCall(CM) & \#  of functions called by a given function\\
\cline{3-4}
& & CountLocalVar (CL) & \#  of local variables in the body of a method\\
\cline{3-4}
& &CountParameters (CPA) & \#  of parameters for a function\\
\cline{3-4}
& & ComToCo (CTC) & the ratio of comments to source code\\
\cline{3-4}
& & CountPath (CP) & the ratio of possible paths in the body of a function\\
\cline{3-4}
& & CyclomaticComplexity (CC) & Mcabe's cyclomatic complexity\\
\cline{3-4}
& & ExecStmt (ES) & \# of executable source code statements\\
\cline{3-4}
& & MaxNesting (MN) & maximum nested depth of all control structures\\
\cline{3-4}
\cline{3-4}
& &{\em Seq. Stat. Mut. Met. (SSMM)} & {\em Description}\\
\cline{3-4}
& &MuS$_{ssdl}$ & \# of generated mutants on $ssdl$ \\
\cline{3-4}
& &MuS$_{swdd}$ & \# of generated mutants on $swdd$ \\
\cline{3-4}
& &MuS$_{oasn}$ & \# of generated mutants on $oasn$ \\
\cline{3-4}
& &MuS$_{oeba}$ & \# of generated mutants on $oeba$ \\
\cline{3-4}
& &MuS$_{olng}$ & \# of generated mutants on $olng$ \\
\cline{3-4}
& &MuS$_{orrn}$ & \# of generated mutants on $orrn$ \\
\cline{3-4}
\cline{2-4}
& &{\em Seq. Dyn. Mut. Met. (SDMM)} & {\em Description}\\
\cline{3-4}
& \multirow{12}{*}{SDyMc} &MuS$_{ssdl}$ & \# of  mutants executed on $ssdl$ \\
\cline{3-4}
& &MuS$_{swdd}$ & \# of mutants executed on $swdd$ \\
\cline{3-4}
& &MuS$_{oasn}$ & \# of mutants executed on $oasn$ \\
\cline{3-4}
& &MuS$_{oeba}$ & \# of mutants executed on $oeba$ \\
\cline{3-4}
& &MuS$_{olng}$ & \# of mutants executed on $olng$ \\
\cline{3-4}
& &MuS$_{orrn}$ & \# of mutants executed on $orrn$ \\
\cline{3-4}
& &MuDuK$_{ssdl}$ & \% of mutants killed $ssdl$\\
\cline{3-4}
& &MuDuK$_{swdd}$ & \% of mutants killed $swdd$\\
\cline{3-4}
& &MuDuK$_{oasn}$ & \% of mutants killed $oasn$\\
\cline{3-4}
& &MuDuK$_{oeba}$ & \% of mutants killed $oeba$\\
\cline{3-4}
& &MuDuK$_{olng}$ & \% of mutants killed $olng$\\
\cline{3-4}
& &MuDuK$_{orrn}$ & \% of mutants killed $orrn$\\
\hline
\end{tabular}
\normalsize
%\vspace*{-12pt}
\end{table*}

\begin{table*}[t]
\centering
\footnotesize
\caption{\label{tab:objects} \textbf{\small Object Program Characteristics}}
%\vspace*{-6pt}
\setlength{\tabcolsep}{3pt}
\begin{tabular}{|l|l|l|l|l|l|l|l|l|c|c|c|} \hline
{\em Program}
& {\em NLOC}
& {\em versions}
& {\em Total inst.}
& {\em Selected insts.}
& {\em Faulty insts.}
& {\em Increase}
& {\em Faults.}
& {\em mutants}
&  {\em tests}
& {\em mutants$_e$}
& {\em mutants$_k$}\\
 \hline
{\sc Apache} & 128K - 201K &  10 (v2.0 - v2.2) & 35,761 &2,164  & 70& 3000\% & 51  & 2,644 &  2,972 &
62\%& 39\% \\
\hline
{\sc Mysql} & 199K - 438K & 10 (v5.0 - v5.6) & 71,665 &2,478 & 230& 1000\% & 184 & 2,267 & 1,813& 71\%
& 33\% \\
\hline
{\sc Mozilla} & 1,120K - 1,268k & 10 (v4 - v34) & 144,382 &4,988  & 142 & 3500\%& 103  & 4,908 & 2,972 &
55\%& 31\% \\
\hline
{\sc OpenOffice} & 3,033K - 4,138K & 10 (v1.0 - v4.1) & 110,509 &6,835  & 116  & 5800\% & 82& 5,097 &
3,846 & 62\%& 32\% \\
\hline
\end{tabular}
\normalsize
\vspace*{-12pt}
\end{table*}

In this section, our goal is to evaluate the effectiveness of \Name{}.
We use four metric suites consisting of 48 metrics in this evaluation, shown in Table~\ref{tab:metrics}.
The CStaMc and CDyMc are static and dynamic metric sets used in  \Name{}.  
In addition to \Name{}, we use SPM, a baseline defect predictor built using 24 metrics,
where SStaMc indicates the static metric set and  SDyMc indicates the dynamic metric set.
We next consider three research questions.

\vspace*{3pt}
\noindent
\textbf {RQ1:
Can \Name{} improve defect prediction performance 
compared to the metrics used for sequential programs?}

\vspace*{3pt}
\noindent
\textbf{RQ2:
What metrics are particularly effective contributors 
to concurrency defect prediction improvement?}

\vspace*{3pt}
\noindent
\textbf{RQ3:
Can a combination of concurrent program metrics be used
to predict concurrency faults of new instances
(i.e., functions) in a new project?
}

\vspace*{3pt}

RQ1 lets us evaluate the defect prediction performance of \Name{},
compared  to the baseline approach in which  code and mutation
metrics for sequential programs are used. 
We also split this overall  question into three sub questions, 
which ask about the effects of static metrics, dynamic metrics,
and machine learning on defect prediction performance. 

\vspace*{3pt}
\noindent
\textbf{RQ1.1:Can applying feature selection improve
the performance of defect prediction?}

\noindent
\textbf {RQ1.2:How effective are the static metrics at 
predicting concurrency faults?}

\noindent
\textbf {RQ1.3:How effective are the dynamic metrics at 
predicting concurrency faults?}

\noindent
\textbf {RQ1.4:How can different machine learners affect
the performance of \Name{}?}

\vspace*{3pt}
\noindent 
RQ2 investigates the contribution of different metrics
for fault prediction because 
choosing a different machine learner can produce 
different performance results.
RQ3 lets us evaluate whether 
the proposed technique is effective when applied
to a new project.

\subsection{Objects of Analysis}
\label{sec:objects}

We study four large concurrent software projects: Apache, MySQL,
Mozilla, and OpenOffice.  We selected these subjects because 
with millions of lines of publicly accessible code and well maintained bug repositories, 
they have been widely used by existing bug characteristic 
studies~\cite{Zaman12,Yin11,Jin12} and
concurrency fault detection and testing 
techniques~\cite{Deng13,Lu07}. In addition, comparing to
medium and small projects, they contain a number of 
concurrency bugs that are more appropriate for 
training datasets.
All four projects started in the early 2000's and each has over 
ten years of bug reports.
The subject programs cover 
various application spectrums - the world's most used HTTP server, the
world's most popular database engine, a leading web browser suite, and
a popular office suite. 
Server applications mostly use concurrency to handle concurrent client
requests. They can have hundreds or thousands of threads 
running at the same time. Client and office applications mostly 
use concurrency to synchronize multiple GUI sessions and background working threads.
Table~\ref{tab:objects} lists our object versions 
along with some of their characteristics. Column~2 lists
the number of lines of non-comment code (NLOC).
Other columns are described later.

%{\sc MySQL}~\cite{mysql} is a free distribution of one of the most widely used
% open source database applications.
%We chose version 5.0.11 ({\sc mysql1}) and version 5.5.3 ({\sc mysql2}).
%We selected these programs because they are representative of real-world code
%% and have been widely used in academic research. In addition, they are each
% applicable to one or more of the
%classes of mutation operators described in Section \ref{background}.

There are not enough concurrency bugs in a single application version
to build classification models (usually 2-5
concurrency bugs per version). Therefore, we collected
data from multiple versions of each application.

We randomly selected 10 versions for each application released
between 2000 and 2014. 
Column~3 of  Table~\ref{tab:objects} lists the number of versions
and release period of each subject.  
Column~4 lists the number of all function instances in all
10 versions of each application.
Since \Name{} targets concurrent programs, we selected function instances
that can be executed by multiple threads. We then removed 
redundant instances across multiple versions. As a result, a total of 16,465
functions (Column 5 in Table~\ref{tab:objects}) were identified for use.

We labeled a function as buggy if it contains at least one concurrency
bug  that was reported in bug reports or release notes (that contain
bug IDs). We searched bug reports for the studied application 
versions using a set of concurrency-related keywords 
(e.g., ``race(s),'' ``deadlock,'' ``atomic,'' ``concurrency," ``synchronization(s),"
``mutex(es)"). These keywords have been used by existing concurrency
bug detection techniques~\cite{Lu08,simracer}.
We filtered out unconfirmed reports.  We then manually
identified functions containing at least one reported bug. 

Column 6 in Table~\ref{tab:objects} lists the number of
faulty function instances. Column 7 lists the number of concurrency bugs.
While having a larger number of bugs may yield better evaluation, 
the cost of the manual process is
quite high: the understanding and preparation of the object 
used in the study and the conduct of the study
required between 150 and 180 hours of researcher time. 
Other columns are described later.

The imbalanced datasets may affect the accuracy of defect prediction~\cite{tan2015online}.
Table~\ref{tab:objects} shows that only 3.4\% of the  instances are buggy and thus the data
is imbalanced. To address this problem and improve defect prediction models, 
we perform the re-sampling technique used in existing work~\cite{tan2015online}, i.e., SMOTE~\cite{chawla2002smote}, on our 
training data for both concurrency and sequential code metric sets.
Column~7 of Table~\ref{tab:objects} shows the percentage of  increase of the minority 
class by the  SMOTE filter.

\subsection{Data collection}

To compute mutation scores, we required mutants of our object programs. 
To seed sequential faults, we use Clang~\cite{llvm} to implement
a mutation generation tool applying the mutation operators
described in Figure~\ref{tab:seq-operators}. 
For concurrency mutants,
we extended {\sc CCmutator}\cite{Kusano13} to create
concurrency mutants of the classes 
described in Section \ref{background}.
This process left us with the numbers 
of mutants reported in Column 9 of Table~\ref{tab:objects}.
A total of 36 mutation metrics, including both
static and dynamic metrics, are collected as shown
in Table~\ref{tab:metrics}.

Test oracles are needed when
evaluating whether a mutant is killed.
These programs are released with existing test suites and with built-in
oracles provided, and we used those.  
We also checked program outputs,
including messages printed on the console and 
files generated and written by 
the programs.

We executed our test cases on all of the 
mutants of each object program. Column 10 of 
Table~\ref{tab:objects} lists the number of test cases. 
To control for variance due to randomization
of thread interleavings, we ran each mutant 100 times.    
A mutant is marked as being executed or propagated
if it does this at least once. 
We used a Linux cluster to perform the executions, distributing each
job on a distinct node.  The mutation score was
computed by following the process described in Section~\ref{ma}.
Columns~11-12 of Table~\ref{tab:objects} report the percentage
of mutants executed and killed. 

To gather static code metric data, for each function 
we first computed six concurrency-related code metrics. 
The method of computing concurrency metrics is described 
in Section~\ref{CPM}. 
%The six sequential metrics include number of 
%instructions (CI), number of basic blocks (CB),
%the number of parameters (CP), the number of function calls
%(CM), and the number of variables (CV).
%McCabe's complexity (CC). These metrics are listed
%in Table~\ref{tab:metrics}.
%All metrics are collected using LLVM pass.
%As for the code coverage metric, 
%we used the open source test coverage tool GCov to measure
%statement coverage (SCV). 
%To measure interleaving coverage (CCV), 
%we followed the approach described in Section~\ref{tm}.
%To record thread information, we used PIN \cite{Luk05} to instrument
%the entry point of each function and recorded thread IDs
%that exercised the function. Column 8-9 of Table~\ref{tab:objects}
%report the statement coverage and interleaving coverage, respectively. 
%

We next computed sequential code metrics (SCM) used as the
baseline approach. 
There are two traditional suites of code metrics: The Chidember-Kemmerer (CK) metrics~\cite{chidamber1994metrics} 
and metrics that are directly calculated at the method level~\cite{subramanyam2003empirical, basili1996validation}. 
The CK metrics measure the size and complexity of 
various aspects of object-oriented source code and 
are calculated at the class level. 
CK metrics have been successfully applied for bug prediction in prior work~\cite{giger2012method}. 
The method level metrics are not limited to object-oriented source code, but include
measures such as lines of code. When applying these metrics to
source files, they are typically averaged or summed up over all methods 
that belong to a file~\cite{lessmann2008benchmarking, nguyen2010studying, zimmermann2008predicting}.
Since our goal is to build fault prediction models for C/C++ programs at 
the function level, we do not use the CK metrics 
because they are not directly applicable to functions, e.g., depth of the inheritance tree. 
We choose instead the nine metrics used previoiusly~\cite{giger2012method}, 
a method-level fault prediction
technique (SCM).  The nine sequential metrics include the number of functions that
call a given function (funIN), the number of functions called by a given function
(funOut), the number of local variables in the body of a method (localVar), 
the number of parameters in the declaration (parameters), the ratio
of comments to source code (comToCo), the number of possible paths
in the body of a function (countPath), McCabe Cyclomatic complexity
of a function (complexity), the number of executable source code
statements (execStmt), and the maximum nested depth of all control structures (maxNesting).

Table~\ref{tab:selection} summarizes all concurrency metrics
and sequential metrics.
The static metric set of \Name{} (i.e., CStaMc) consists
of a static code metric set, denoted by CCM, 
and a static mutation metric set, denoted by CSMM. 
The dynamic metric set of \Name{} (i.e., CDyMc) is equal to
the dynamic mutation set, denoted by CDMM. 
On SPM, the static metric set (i.e., SStaMc) consists of
a code metric set (SCM) and a mutation metric set (SSMM) specific to 
sequential programs. SPM's dynamic metric set uses a set of
dynamic mutation metrics for sequential programs, denoted by SDMM.
The sequential code metric set 
SCM is proposed in prior work~\cite{giger2012method} and used 
as a baseline approach in our study.

\subsection{Techniques for Comparison}
\label{sec:techniques}
The \Name{} basically combines all proposed static and dynamic concurrency
metrics, i.e., CCM + CSMM + CDMM. To answer RQ1, we compare \Name{}
to SPM (i.e., SCM + SSMM + SDMM).
The subquestion RQ1.1 isolates CStaMC from \Name{} and compares it 
to \Name{}, SStaMC, and CDyMC.
The subquestion RQ1.2 compares CDyMc 
to \Name{}, and SDyMC.
To answer the subquestion RQ1.3,  we 
compare the results of prediction models using four different 
classification algorithms widely adopted in defect
prediction studies,  including Decision Tree, Logistic
Regression, Naive Bayesian, and Random Forest.
To answer RQ2, we calculate the importance  
of each individual metric used in \Name{} and 
examine their contribution in concurrency defect prediction. 
To answer RQ3, we apply the model learned from 
each dataset  to predict concurrency faults
in each of the other three datasets. We then evaluate
the prediction performance of both 
\Name{} and SPM on all nine pairs of comparison.

\subsection{Prediction Models}
\label{subsec:vars}

We first performed feature selection to select effective metrics
for use in constructing prediction models. To do this, 
we used the WEKA Wrapper Subset Selection 
Filter~\cite{kohavi1997wrappers,hall2009weka} ({\tt WrapperSubsetEval}), which performs a best first
search algorithm to identify the subset of attributes that generalize 
best on the training set.
The SS algorithm can resolve the multicollinearity
problem between correlated features~\cite{Alpaydin04}
and thereby avoid the model construction overfitting problem. 
The {\tt WrapperSubsetEval} evaluates the power of a subset of metrics by
considering the individual predictive ability of each metric along with the degree of 
redundancy between them. Metrics that are highly correlated 
with the faulty class while having a low inter-correlation are preferred.

\begin{table}[t]
\centering
\footnotesize
\caption{\label{tab:selection} \textbf{\small Metrics selected
in each metric suite for all subjects}}
%\vspace*{-6pt}
\setlength{\tabcolsep}{1pt}
\begin{tabular}{|p{3cm}|p{6cm}|} \hline
{\em Concurrency Metric Suite}
& {\em Selected metrics}\\
 \hline
\Name{}  & 
CSV, CCC, CCD, CCE, CSO, MuS$_{rmlock}$,
MuDuE$_{spltecs}$, MuDuE$_{rmsig}$,
MuDuE$_{swptw}$, MuDuE$_{mwait}$, 
MuDuK$_{swptw}$, MuDuK$_{rmsig}$, MuDuk$_{rmlock}$
\\
\hline
CStaMc & 
CSM, CCC, CCD, MuS$_{rmlock}$,
MuS$_{swptw}$, MuS$_{rmsig}$. 
\\ 
\hline
CDyMc & 
MuDuE$_{rmsig}$, MuDuE$_{rmlock}$,
MuDuE$_{swptw}$, MuDuE$_{mwait}$, 
MuDuE$_{spltecs}$,
MuDuK$_{swptw}$, MuDuK$_{rmsig}$
\\
\hline
\hline
{\em Sequential Metric Suite}
& {\em Selected metrics}\\
\hline
SPM & CI, CP, CC,
MuS$_{ssdl}$, MuS$_{oasn}$, MuS$_{orrn}$,
MuDuE$_{swdd}$, MuDuE$_{oasn}$,
MuDuK$_{ssdl}$, MuDuK$_{oeba}$
\\
\hline
SStaMc & CI, CP, CC, CV,
MuS$_{ssdl}$, MuS$_{olng}$, MuS$_{swdd}$
\\
\hline
SDyMc  & MuDuE$_{swdd}$, MuDuE$_{oasn}$,
MuDuE$_{ssdl}$,
MuDuK$_{ssdl}$, MuDuK$_{oeba}$,
MuDuK$_{olng}$\\
%\hline
%SMuMc (SSMM + SDMM) & ?\\
\hline
\end{tabular}
\normalsize
\vspace*{-12pt}
\end{table}

Next, a classification algorithm was required to build the prediction model for
each subject. In \Name{},  we consider four classification techniques:
Bayesian Network, J48 Decision Tree, 
Logistic Regression, and  Random Forest in Weka~\cite{Hall09}. 
We chose them because they are popular and have been 
shown to be effective at predicting defects in a recent study~\cite{Ghotra2015}.
Naive Bayes (NB)~\cite{witten2005data} is a statistical technique which uses the
combined probabilities of the different attributes to predict faultiness. 
Logistic Regression (LR)~\cite{cox1958regression} is a regression technique
which identifies the best set of weights for each attribute to predict the 
faulty or non-faulty class. 
J48 is a Java implementation of the C4.5~\cite{quinlan2014c4} decision tree
algorithm which uses entropy information to determine which attribute to use as decision nodes. 
Random Forest (RF)~\cite{breiman2001random}  is an ensemble technique which
aggregates the predictions made by a collection of decision trees (each with a subset of the original set of attributes).
To examine our research questions, we 
applied the four models to different metric sets. 
The random forest algorithm was primarily used in
our experiments because its performance was good, 
as noted in Section~\ref{subsub:rq1.3}. 

To evaluate our prediction models, we again used 10-fold cross validation, 
widely used to evaluate prediction models~\cite{Lee11,Moser08}. 

In 10-fold cross validation we randomly divide the dataset into ten folds. Of these ten folds, nine folds are used to train the classifier, while the remaining one fold is used to evaluate the performance
(Section~\ref{subsec:performance}).  The feature
selection is performed on the training set. Specifically, 
the {\tt WrapperSubsetEval} selected the best features
(metrics) in a fold. Next, 
only the metrics that were nominated were adopted in the model construction. 
This selection and model construction process was iterated  for each of the ten folds. 
%only the metrics that were nominated at least more than twice 
%(in two different folds) were finally adopted in the model construction.
%a classification algorithm was required to build the 
%prediction model for each of the four datasets (subjects).
Table~\ref{tab:selection} shows the selected metrics for each metrics suite in 
all subjects.

Since 10-fold cross validation randomly samples instances and puts them in ten folds~\cite{Alpaydin04},
we repeated this process 100 times for each prediction model to avoid 
sampling bias~\cite{Lee11}. Note that we use Weka's SMOTE filter to increase 
the instances of the minority class. The filter is applied only to the training folds
of the cross validation instead of the whole dataset in advance. 
This is because the latter approach is likely to provide over optimistic 
results~\cite{rodriguez2014preliminary}.

\ignore{
In addition to the three classification techniques,
we use ZeroR as the baseline
for comparison with our approach.  
During training, ZeroR ignores the features 
and relies only on the labels for predicting. 
Although it does not have much predicting capability, it establishes 
the lowest possible predictability that a classifier should have. 
It works by constructing a frequency table for the labels in the 
training data and selects the most frequent values of the testing 
data in predicting.
}

\subsection{Performance Metrics}
\label{subsec:performance}

We chose performance metrics allowing us
to answer each of our three research questions.
Specifically, we employ precision, recall, 
F1-measure,  area under the curve (AUC), and
cost-effectiveness metric. 
An instance can be classified as: buggy
when it is truly buggy (true positive, TP); 
it can be classified as buggy when it is actually not (false positive, FP); 
it can be classified as non-buggy when it is actually buggy (false negative, FN); 
or it can be correctly classified as non-buggy (true negative, TN).

\subB{Precision,  Recall, and F1-measure.}

\begin{itemize}

\item\textbf{Precision:} the number of instances correctly classified 
as buggy over the number of 
all instances classified as buggy.

\begin{center}
$P$ = $\frac{TP}{TP + FP}$
\end{center}

\item\textbf{Recall:} the number of instances correctly classified as
buggy over the total number of buggy instances.

\begin{center}
$R$ = $\frac{TP}{TP + FN}$
\end{center}

\item\textbf{F-measure:} a composite measure of precision and
recall for buggy instances.

\begin{center}
$F(b)$ = $\frac{2*P*R}{P + R}$
\end{center}

\end{itemize}

\subB{AUC (area under the curve).}
We use the AUC of the receiver operating characteristics (ROC)~\cite{lessmann2008benchmarking} 
as an additional measure to evaluate the performance of the prediction models.  
The range of AUC is [0, 1]. A larger AUC score indicates better prediction performance. 
A prediction model achieving AUC above 0.5 is considered more effective than the random predictor
and 0.7 is reasonably good~\cite{hosmer2013applied}.
As a scalar value, AUC is well suited to compare the performance of different classifiers, 
and is often used for that purpose~\cite{mende2009revisiting}.

\subB{Cost Effectiveness.}
The cost effectiveness metric, which evaluates prediction performance given a cost limit,
 has been  used widely in 
existing defect prediction techniques~\cite{yang2016effort,
kamei2013large, xia2016hydra, lessmann2008benchmarking}.
In our context, the cost is the amount of function instances to inspect, 
and the benefit is the number of bugs that can be discovered. 
If developers inspect all predicted buggy instances, the percentage of 
bugs that can be detected is equivalent to the recall. 
In some circumstances (e.g., meeting a deadline), developers can
only inspect certain amount of functions. Therefore, it is useful 
to maximize the bugs to be detected while minimizing the number of 
instances to inspect. In this case, the cost effectiveness metric is appropriate.
We use the cost effectiveness metric, PofB20,  used by Jiang et al.~\cite{jiang2013personalized}. 
They measure the percentage of bugs that a developer can identify by inspecting the top 20 
percent of lines of code. In our context, we measure the percentage of concurrency faults
that a developer can identify by inspecting the top 20  percent of function instances. 

To compute PofB20,  we sort instances by their probability (provided by WEKA) of  buggy~\cite{rahman2013and}
We then simulate a developer that inspects these potentially buggy instances one at a time.
As the instances are inspected one at a time, we count the number of lines of code that have been inspected 
and the number of bugs that have been identified. 
We stop the inspection process when 20 percent of the lines of code have been inspected and
compute the percentage of bugs that are identified. 
This percentage is the PofB20 cost effectiveness score. A higher cost effectiveness 
score represents  that a developer can detect more bugs when 
inspecting a limited number of lines of code.

PofB20 metric uses 20 percent of all effort as the cut-off value. However, 
a different cut-off value might lead to different results.  As an additional metric,
we use $P_{opt}$~\cite{mende2009revisiting} to evaluate the prediction 
performance of models. $P_{opt}$ is defined as the area  $\Delta_{opt}$ between 
the optimal model and the prediction model.
In the optimal model, all instances are ordered by  decreasing fault density, 
and in the predicted model, all instances are ordered by decreasing predicted value
(i.e., probability of being buggy). 
The equation of computing $P_{opt}$ is shown below, where a 
larger $P_{opt}$ value means a smaller difference between the 
optimal and predicted model:

\begin{center}
$P_{opt}$ = $1 - \Delta_{opt}$
\end{center}

\noindent
The range of $P_{opt}$ is [0, 1]  and any predictor achieving the $P_{opt}$ above 0.5 is more effective than the random predictor.

\subB{Statistical significance analysis.}
To assess whether prediction performance of different
metric sets were statistically significant, we applied the Wilcoxon
test~\cite{walpole1993probability} to the data sets, comparing each pair of metric sets within each model.
We did not use t-test because F-measure outcomes from cross validation did not
follow a Normal distribution.
 We checked if the mean of F-measure values of one predictor $P_i$ was greater than
 the mean of F-measures of another predictor $P_j$  at the 95\% confidence level
 ($p-value < 0.05$).

Specifically, the null and alternative hypotheses for the t-test are:

\begin{itemize}
\item \textbf{H0:} 
F-measure mean of $P_i$ is equal to the F-measure mean of $P_j$.

\item \textbf{H1:} F-measure mean of $P_i$ is greater to the F-measure 
mean of $P_j$. (i.e., $P_i$ has better performance if the mean value is higher). 
\end{itemize}

We rejected the null hypothesis H0 and accepted the alternative hypothesis H1 
if the p-value was smaller than 0.05 (at the 95\% confidence level).

\subB{Effect size.}
The effect size uses Cliff's delta~\cite{cliff1993dominance}
that quantifies the amount of difference between two 
non-parametric variables beyond the p-value interpretation.
The Cliff's delta is computed by $d = 2W/mn - 1$,
where $W$ is the statistic of  
the Wilcoxon rank-sum test, and $m$ and $n$ are 
the sizes of two compared distributions.
Here $W$ = $R - n(n+1)/2$, where $R$ is the sum of the 
rank in the sample and $n$ is the sample size.
The magnitude of effect size is usually assessed using the 
thresholds~\cite{romano2006appropriate}, where
$|d|$ $<$ 0.147 is negligible, 0.147 $\geq$ $|d|$ $<$ 0.33 is
small, 0.33 $\geq$ $|d| <$ 0.474 is medium, and otherwise large. 
For example, suppose the effect size
between two metric suites A and B is $-$0.75. The sign is negative
because the mean of A is greater than the mean of B and the magnitude 
of the effect size is regarded as large.

\subsection{Importance of Features}
To study the most influential metrics, we 
compute Breiman's variable importance score~\cite{breiman2001random}
for each feature. The larger the score, the greater the influence of the
metric on our models.  We use the option {\tt -attribute-importance}
provided by Weka to compute the variable importance scores. 
For each run of the 10-fold cross validation we obtain
an importance score  for each feature. 
In order to determine the features that are most influential 
for the whole dataset, we apply the Scott-Knott test \cite{jelihovschi2014scottknott} on 
the values from all 10 runs.  
The Scott-Knott test will cluster the metrics according to statistically significant differences in their mean variable importance scores (p - value = 0.05). We use the implementation of the Scott-Knott test provided by the ScottKnott R package. The Scott-Knott test ranks each metric exactly once, however several metrics may appear within one rank.

Next, to assess  how each factor is related to buggy instances,
we compare the values of each feature between buggy instances
and non-buggy instances. We first analyze the statistical significance of 
the difference between the two classes (buggy and non-buggy) by applying the 
Mann-Whitney U test at p - value = 0.01. 
To show the effect size of the difference between the two features 
in two groups, we calculate Cliff's Delta.

\subsection{Threats to Validity}
\label{sec:threats}

The primary threat to external validity for this
study involves the representativeness of our programs,
mutants, coverage criteria, and test cases.
Other systems may exhibit different behaviors, as may other forms of test cases.
However, the programs we
investigated are popular open source programs.
Furthermore, the test cases are  those provided with the programs: 
they are representative of test cases that could be used in practice to test these programs.
Most of the test subjects we
used had relatively good test suites (i.e., of the covered mutants,
the mutation scores were above 80\%). 
Mutants can be influenced by external factors such as mutation operators.
We used only concurrency mutation operators. However, concurrency faults can also be introduced by sequential glitches.
In addition, other interleaving criteria (e.g., synchronization coverage) may lead to different
coverage results.
We controlled for these threats by using well studied
concurrency mutation operators and popular interleaving criteria.

The data collected for defect prediction may contain noise, such as false positives
(identifying non-buggy changes/files as buggy), random sampling,
and imbalanced data~\cite{kim2011dealing, bird2009fair}. 
For example,
Bird et al.~\cite{bird2009fair} discovered that data collected via automated mining software
repositories (MSR) often contain noise.  To mitigate this threat, 
we manually and carefully selected high quality datasets. The bugs we selected
are confirmed and fixed in the subsequent versions. While it is possible 
that some functions contain concurrency bugs but were mislabeled  as clean,
we selected program versions between 2000 and 2014, so the functions labeled as clean 
are unlikely to contain concurrency bugs because no such bugs have been 
reported since 2014.
We use SMOTE to handle the class imbalance problem.

The primary threats to internal validity for this study
are possible faults in the implementation of our approach
and in the tools that we used to perform evaluation.
We controlled for this threat by extensively testing
our tools and verifying their results against smaller
programs for which we could manually determine the correct results.
We also chose to use popular and established tools (e.g., LLVM, Weka)
for implementing the various stages of our approach.

Where construct validity is concerned, our measurements
involve using metrics extracted from source code and 
mutation analysis to predict
defects in concurrent programs.
Other static metrics and dynamic metrics (e.g., test suite
metrics) are also of interest.
Furthermore, other machine learning performance measures
can be used to measure effectiveness and accuracy.
To control for this threat, we chose commonly used F-measures.
Other metrics, such as the Matthews Correlation
Coefficient~\cite{shepperd2014researcher} (MCC),  
can be used to handle unbalanced data.

Conclusion validity concerns the statistical significance of the result. 
We applied 10-fold cross validation, and did so 100 times, as is common in 
experiments of this type.  We also undertook statistical analysis to test our hypotheses.  
To further reduce threats to conclusion validity, we were careful to check the 
assumptions of the statistical tests that were used.

%http://masterr.org/r/an-easy-way-to-make-ggplot2-histograms-ezplot-part3/
%http://chrisalbon.com/r-stats/histograms-and-density-plots.html
%http://www.statmethods.net/graphs/density.html

%install.packages("devtools")
%devtools::install_github("gmlang/ezplot")
%library(ezplot)
%plt = mk_distplot(iris)
%title1 = "Histogram of Sepal Length"
%p = plt("Sepal.Length", main=title1)
%print(p)
%p = plt("Sepal.Length", type="density", main=title1, add_vline_mean=T)

\ignore{
\begin{figure*}[t!]
  \begin{center}
\includegraphics[scale=0.5]{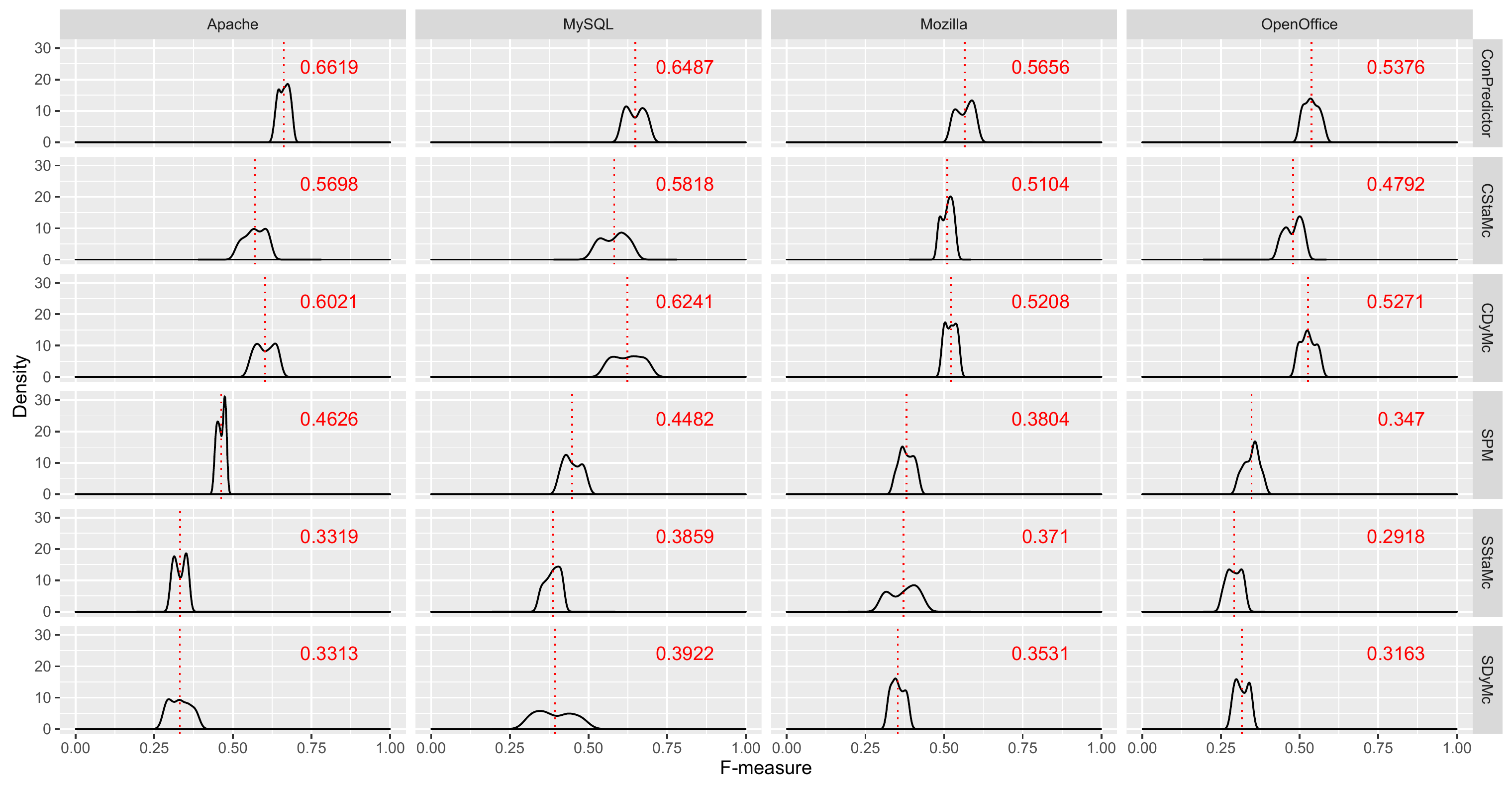}
  \end{center}
  \caption{\textbf{\small Performance comparison of prediction models by different subjects.}}
  \label{fig:performance}
% \vspace{-12pt}
\end{figure*}
}

\section{Results and Analysis}
\label{results}

In this section, we present results related to the three research questions.
\footnote{\scriptsize All data we used in our experiments are publicly
 available at
\url{http://cs.uky.edu/~tyu/ConPredictor}}

\subsection{RQ1: Effectiveness of \Name{} vs. SPM vs. SCM} 
To examine RQ1, we compare the performance of 
\Name{} to that of SPM and SCM.
%We use F-measure to evaluate prediction performance for the three classification
%algorithms, as described in Section~\ref{study}.
 Columns 3-5 and 7-9 of Table~\ref{tab:metrics} show the performance of each set of metrics
in different subjects 
in terms of precision, recall, F-measure, AUC, PofB20,  and P$_{opt}$
from 100 times ten-fold 
cross validations. The numbers marked in boldface indicate that 
the performance of \Name{} is significantly different from 
that of SPM and SCM.  Columns 6 reports the F-measure values 
of the techniques without applying feature selection.
%We gained 100 F-measures from the repeated cross validation
%process and plot the density plots. The text on a subfigure
%indicates the mean  value across all 100 F-measures. 

Although the performance values varied, there was 
a clear trend in which \Name{} 
outperformed SPM and SCM for every subject. 
For example, comparing to SPM, the improvement of F-measure ranged
from 29.7\% to 34\%. Comparing to SCM, 
the improvement of F-measure ranged
from 37.7\% to 50\%. 
The most improvement occurred on 
Apache while the least improvement was seen with OpenOffice.  
In other words, metrics considering concurrency characteristics
improved the prediction performance. 

Table~\ref{tab:cc} shows  the effect size when 
comparing different metric suites in terms of their F-measure values. 
The numbers marked as bold indicate that the Wilcoxon rank-sum test 
rejected the null hypothesis of RQ1 (p-value $<$ 0.05).
For example, on Apache, the effect size between
\Name{} and SPM is $\minus$0.675 and they are 
{\em statistically} significantly different. 
Comparing \Name{} to SPM,
the largest effect size occurred with
OpenOffice and smallest
effect size occurred for Mozilla.

Among all 558 buggy instances,  368 and 270 are predicted 
as buggy by \Name{} and SPM, respectively.
All 270 faults predicted by SPM are also predicted by 
\Name{}. 
SPM has a lower precision because
the missing buggy instances typically have lower sequential metric
scores but higher concurrency metric scores. For example, 
a function that does not contain any branches (CyclomaticComplexity
is 2) but has several unprotected shared variables involves
an atomicity violation bug but is falsely classified as non-buggy by
SPM. 

In all four subjects, the AUC values of \Name{} are above
0.7 and significantly better than SPM and SCM. 
This result confirms the impact of our proposed concurrency
metrics on concurrency fault prediction. 

PofB20 represents the number of bugs that can be discovered by 
examining the top 20\% LOC. 
For example, \Name{} can help the developers identify 24 bugs for Apache by inspecting 20\% LOC. 
SCM can help identify 12 bugs by inspecting 20\% LOC, 
which is 12 less bugs than those of \Name{}. 
Overall,  \Name{} improved PofB20 over SPM by amounts  raninge from
5.8\% to 23.4\% and over SCM by amounts ranging from 30.8\% to 53.8\%.

Overall, these results  
suggest that {\em
\Name{} outperforms the  traditional sequential metrics}.

\begin{table*}[t]
\centering
\small
\caption{\label{tab:metrics} \textbf{\small Results of evaluated metrics}}
%\vspace*{-6pt}
\setlength{\tabcolsep}{10pt}
\begin{tabular}{|l|l|l|l|l|l|l|l|l|} \hline
Project & Technique & P & R & F1 &F1$_{nf}$ & AUC & PofB20 & $P_{opt}$  \\
\hline
\hline
\multirow{6}{*}{Apache} & ConPredictor & \textbf{0.68}& \textbf{0.64}& \textbf{0.66} & 0.62$\downarrow$& \textbf{0.72} & \textbf{0.52} & \textbf{0.67}\\
\cline{2-9}
& CStaMc& 0.65 & 0.51 & 0.57 &  0.51$\downarrow$ & 0.68 & 0.51  & 0.61   \\
\cline{2-9}
& CDyMc& 0.64 &  0.56 & 0.60  & 0.55$\downarrow$ & 0.52 & 0.54 & 0.50 \\
\cline{2-9}
& SPM& 0.52 &0.41 & 0.46 & 0.42$\downarrow$ &  0.48 & 0.49 & 0.44 \\
\cline{2-9}
& SStaMc& 0.43 & 0.31 & 0.33 & 0.31 & 0.44 & 0.38 & 0.42  \\
\cline{2-9}
& SDyMc& 0.46 & 0.32 &0.35 & 0.29$\downarrow$  & 0.47 & 0.34 & 0.44  \\
\cline{2-9}
& SCM& 0.43 & 0.34 & 0.33 & 0.31 & 0.45 & 0.36 & 0.41  \\
\hline
\hline
\multirow{6}{*}{MySQL} & ConPredictor & \textbf{0.65} & \textbf{0.63} & \textbf{0.64} &  0.55$\downarrow$& \textbf{0.74}  & \textbf{0.42} & \textbf{0.69} \\
\cline{2-9}
& CStaMc& 0.59 & 0.57& 0.58 & 0.52$\downarrow$ &  0.64 & 0.40 & 0.58 \\
\cline{2-9}
& CDyMc& 0.61 & 0.63 &0.62 & 0.56$\downarrow$ & 0.66 & 0.42 & 0.60 \\
\cline{2-9}
& SPM& 0.52 & 0.40 &0.45 & 0.41 & 0.51 & 0.35 & 0.51 \\
\cline{2-9}
& SStaMc& 0.35 & 0.38 & 0.41 & 0.35$\downarrow$ & 0.48 & 0.31 & 0.50 \\
\cline{2-9}
& SDyMc& 0.38 & 0.33 & 0.39 &0.32$\downarrow$ & 0.48 & 0.28 & 0.45 \\
\cline{2-9}
& SCM& 0.32 &0.32  & 0.36 &0.32$\downarrow$ & 0.47 & 0.25 & 0.42  \\
\hline
\hline
\multirow{6}{*}{Mozilla} & ConPredictor & \textbf{0.60} & \textbf{0.54} & \textbf{0.57} & 0.52$\downarrow$ & \textbf{0.75} & \textbf{0.47} &\textbf{0.70} \\
\cline{2-9}
& CStaMc& 0.55 & 0.48 &0.51 & 0.45$\downarrow$ & 0.55 & 0.39 & 0.52 \\
\cline{2-9}
& CDyMc& 0.57 & 0.48&0.52 & 0.47$\downarrow$ &  0.58 & 0.42 &  0.55 \\
\cline{2-9}
& SPM& 0.54 & 0.29 & 0.38 & 0.35 & 0.44 & 0.36 & 0.54 \\
\cline{2-9}
& SStaMc& 0.41 & 0.34 & 0.37 & 0.35 &  0.40 & 0.36 & 0.38 \\
\cline{2-9}
& SDyMc& 0.42 & 0.30 & 0.35 & 0.32 & 0.44 & 0.34 & 0.40 \\
\cline{2-9}
& SCM& 0.41 & 0.31 & 0.35 & 0.33$\downarrow$ & 0.42 & 0.32 & 0.42 \\
\hline
\hline
\multirow{6}{*}{OpenOffice} & ConPredictor & \textbf{0.59} & \textbf{0.48} & \textbf{0.53} & 0.50$\downarrow$ & \textbf{0.77} & \textbf{0.52} & \textbf{0.68} \\
\cline{2-9}
& CStaMc& 0.57 & 0.41 &  0.48 & 0.46$\downarrow$ & 0.56 & 0.50 & 0.52\\
\cline{2-9}
& CDyMc& 0.60 & 0.47 & 0.53 & 0.51$\downarrow$ & 0.58  & 0.51 & 0.53 \\
\cline{2-9}
& SPM& 0.48& 0.28 & 0.35 & 0.33$\downarrow$ & 0.46 & 0.40 & 0.42 \\
\cline{2-9}
& SStaMc& 0.35 & 0.25 &0.29  & 0.28 & 0.44 & 0.36 & 0.42 \\
\cline{2-9}
& SDyMc& 0.39 &0.27 &0.32 & 0.29$\downarrow$ & 0.46 & 0.24 &  0.41\\
\cline{2-9}
& SCM& 0.37 & 0.30 & 0.33 & 0.31 & 0.42 &0.24 & 0.39 \\
%\hline
%\hline
%\multicolumn{2}{|c|}{Average}  & ? & & 0.53& & & &\\
\hline
\end{tabular}
\normalsize
%\vspace*{-12pt}
\end{table*}

% Refer to Jiang et al. paper. 

\begin{table*}[t]
\centering
\small
%\addtolength{\tabcolsep}{-3pt}
\caption{\label{tab:cc} \textbf{\small  Effect Sizes for
each metric set in different subjects. }}
%\vspace*{-6pt}
%\setlength{\tabcolsep}{3pt}
\begin{tabular}{|l||c|c|c|c|c|c||c|c|c|c|c|c|} \hline
{\em Prog.}  
& \multicolumn{6}{c||} {\sc ConfPredictor}
& \multicolumn{6}{c|} {\sc CStaMc}
\\ 
& CStaMc & CDyMc & SPM & SStaMc & SDyMc &SCM
& CStaMc & CDyMc & SPM & SStaMc & SDyMc &SCM\\ \hline
Apache & \textbf{-0.223} & \textbf{-0.142} & 
\textbf{-0.675} & \textbf{-0.734} & \textbf{-0.539} & 
\textbf{-0.777} & - & 0.133 & \textbf{-0.424} &  \textbf{-0.498} &  \textbf{-0.432}  &   \textbf{-0.501} \\
\hline
MySQL & \textbf{-0.374} & \textbf{-0.258} & 
\textbf{-0.842} & \textbf{-0.577} & \textbf{-0.699} & 
\textbf{-0.850} & - &\textbf{0.241} & \textbf{-0.692} &  \textbf{-0.701} &  \textbf{-0.522} &  \textbf{-0.725} \\
\hline
Mozilla & \textbf{-0.331} & \textbf{-0.251} & 
\textbf{-0.603} & \textbf{-0.658} & \textbf{-0.694} & 
\textbf{-0.709}& - &0.231 & \textbf{-0.498} &  \textbf{-0.572} &  \textbf{-0.451} &  \textbf{-0.533}\\
\hline
OpenOffice & \textbf{-0.402} & \textbf{-0.333} & 
\textbf{-0.862} & \textbf{-0.801} & \textbf{-0.792} & 
\textbf{-0.877} & - &\textbf{0.312} & \textbf{-0.398} &  \textbf{-0.402} &  \textbf{-0.257} &  \textbf{-0.429}\\
\hline
\end{tabular}

\vspace{5pt}

\begin{tabular}{|l||c|c|c|c|c|c|}
\hline
{\em Prog.}  & \multicolumn{6}{c|} {\sc CDyMc}
\\ 
& CStaMc & CDyMc & SPM & SStaMc & SDyMc &SCM\\ \hline
Apache &
 - & - & \textbf{-0.563} &  \textbf{-0.852} & \textbf{-0.771} &  \textbf{-0.642} \\
\hline
MySQL & 
- & - & \textbf{-0.701} &  \textbf{-0.721} & \textbf{-0.814} & \textbf{-0.759} \\
\hline
Mozilla & 
- & - & \textbf{-0.423} &  \textbf{-0.499} & \textbf{-0.367} & \textbf{-0.598} \\
\hline
OpenOffice &
- & - & \textbf{-0.598} &  \textbf{-0.744} & \textbf{-0.725} & \textbf{-0.415} \\
\hline
\end{tabular}
 %\begin{tablenotes}
 %     \small
 %     \item The effect sizes when comparing two techniques. The
 %     bolded numbers indicate that the two techniques are statistically
 %     different.
 %   \end{tablenotes}
\normalsize
%\vspace*{-5pt}
\end{table*}

\begin{table}[t]
\centering
\small
\caption{\label{tab:ml} \textbf{\small Performance comparison of prediction models by different machine learners.}}
%\vspace*{-6pt}
%\setlength{\tabcolsep}{1pt}
\begin{tabular}{|l|l|l|l|l|} \hline
Learner & B.N & D.T. & L.R. & R.F.  \\
\hline
\hline
ConPredictor & 0.57 & 0.61 & 0.53 & 0.65 \\
\hline
CStaMc& 0.48 & 0.47 & 0.47  & 0.57  \\
\hline
CDyMc& 0.53 & 0.55 & 0.49  & 0.62 \\
\hline
SPM &0.39 & 0.40 & 0.37 &  0.45\\
\hline
 SStaMc &0.34 & 0.33 & 0.32 & 0.38 \\
\hline
SDyMc& 0.37 & 0.36 & 0.34 & 0.40 \\
\hline
SCM& 0.32 & 0. 31 & 0.32 & 0.35\\
\hline
\end{tabular}
\normalsize
%\vspace*{-12pt}
\end{table}

\begin{table*}[t]
\centering
\small
%\addtolength{\tabcolsep}{-6pt}
\caption{\label{tab:ml} \textbf{\small  Effect sizes for
each machine learner in different subjects. }}
%\vspace*{-6pt}
%\setlength{\tabcolsep}{3pt}
\begin{tabular}{|l||c|c|c||c|c|c||c|c|c|} \hline
{\em Learner}  
& \multicolumn{3}{c||} {\sc ConfPredictor}
& \multicolumn{3}{c||} {\sc CStaMc}
& \multicolumn{3}{c|} {\sc CDyMc}
\\ 
& B.N. & D.T. & L.R. 
& B.N. & D.T. & L.R.  
& B.N. & D.T. & L.R.  \\ \hline
B.N. & - & 0.122 & \textbf{-0.358} & 
- & -0.025 & -0.152 & 
0.231 & -0.101 & -0. 132\\
\hline
D.T.  & \textbf{-0.544} &  - & \textbf{0.205} & 
-0.332 & - & 0.108 & 
0.098 &  -  &  0.133\\
\hline
L.R.  & \textbf{0.198} & 0.330 & - & 
0.182 & \textbf{0.258}  &  - & 
\textbf{0.209} & \textbf{0.298} & -\\
\hline
R.F. & \textbf{-0.499} & 0.223  & \textbf{-0.532} & 
\textbf{-0.392} &  \textbf{-0.254} & \textbf{-0.338} & 
 \textbf{-0.292}& \textbf{-1.163}   & \textbf{-0.083}\\
\hline
\end{tabular}
\normalsize
\vspace*{-6pt}
\end{table*}

\begin{table}[t]
\centering
\small
\caption{\label{tab:rank} \textbf{\small Scott-Knott test results.}}
%\vspace*{-6pt}
\setlength{\tabcolsep}{4 pt}
\begin{tabular}{|l|l|l|l|l|} \hline
Group & Metric & Rank$_{mean}$ & Rank$_{highest}$ & Rank$_{lowest}$  \\
\hline
\hline
\multirow{3}{*}{G1} & CCC  & 1 & 1 & 1 \\
\cline{2-5}
&MuDuE$_{spltecs}$  & 1.6 & 1 &2\\
\cline{2-5}
& MuDuS$_{rmlock}$ & 1.8  & 1 & 3\\
\hline
\multirow{2}{*}{G2} & MuDuK$_{rmlock}$  &  3.1  & 2 & 5\\
\cline{2-5}
& CSV & 4.9 & 3 & 6 \\
\hline
\multirow{3}{*}{G3} & MuDuK$_{swptw}$ &  5.8 &  4 & 7\\
\cline{2-5}
& MuDuE$_{mwait}$ &6.2 & 5 &  7\\
\cline{2-5}
& MuDuE$_{swptw}$ & 7.5 & 6  & 9\\
\hline
\multirow{2}{*}{G4} &  CCE & 9.1 &  8  &  10\\
\cline{2-5}
& MuDuK$_{rmsig}$ & 10.6 & 9  & 11\\
\hline
\multirow{1}{*}{G5} & CCD & 10.9 & 10  & 12 \\
\hline
\multirow{1}{*}{G6} & CSO & 11.5 & 10 & 13\\
\hline
\multirow{1}{*}{G7} & MuDuE$_{rmsig}$& 12.2 & 11 & 13 \\
\hline
\end{tabular}
\normalsize
%\vspace*{-12pt}
\end{table}

\begin{table}[t]
\centering
\small
\caption{\label{tab:effects} \textbf{\small Effects of different metrics}}
%\vspace*{-6pt}
\setlength{\tabcolsep}{10pt}
\begin{tabular}{|l|l|l|l|} \hline
Metric set & Metric & Rel. & d-value  \\
\hline
\hline
\multirow{3}{*}{CStaMc} & CCC & +  & \textbf{0.341 (medium)} \\
\cline{2-4}
& CSV & + &  0.225 (small)\\
\cline{2-4}
& MuDuS$_{rmlock}$  & + &  0.204 (small)\\
\hline
\hline
\multirow{4}{*}{CDyMc} & MuDuE$_{rmlock}$ &  + &  \textbf{0.325 (medium)} \\
\cline{2-4}
& MuDuE$_{spltecs}$& + & 0.212 (small)\\
\cline{2-4}
& MuDuk$_{rmlock}$& + &   \textbf{0.308 (medium)}\\
\cline{2-4}
& MuDuK$_{swptw}$ & + & 0.187 (small)\\
\hline
\end{tabular}
\begin{flushleft}
Rel.= relationship. ``+" indicates faulty instances  have significantly higher value on this metric. 
\end{flushleft}
\normalsize
%\vspace*{-12pt}
\end{table}

\begin{table*}[t]
\centering
\small
\caption{\label{tab:cp} \textbf{\small  F-measures on across project prediction
using \Name{} and SPM.}}
%\vspace*{-6pt}
%\setlength{\tabcolsep}{3pt}
\begin{tabular}{|l|c|c|c|c||c|c|c|c|} \hline
{\em Prog.}  
& \multicolumn{4}{c||} {\sc ConfPredictor}
& \multicolumn{4}{c|} {\sc SPM}
\\ 
& Apache
& MySQL
& Mozilla
& OpenOffice
& Apache
& MySQL
& Mozilla
& OpenOffice\\ 
\hline
Apache  & - & \textbf{0.546} &  \textbf{0.566} &  \textbf{0.496} & - & 0.421 & 0.443 & 0.402\\
\hline
MySQL & 0.311 & - &  \textbf{0.399} &  \textbf{0.475} & 0.294 & -   & 0.315 & 0.358\\
\hline
Mozilla &  \textbf{0.502} &  \textbf{0.387} & - &  0.375 & 0.394 & 0.302 & - & 0.331\\
\hline
OpenOffice &  \textbf{0.451} &  0.325 &  \textbf{0.364} & - & 0.425  & 0.308 & 0.341 & -\\
\hline
\end{tabular}
\normalsize
\vspace*{-12pt}
\end{table*}

\subsubsection{RQ1.1: Feature Selection}

As Column 6 of Table~\ref{tab:metrics} shows, 
feature selection improves performance of
the defect prediction in terms of F-measure 
in all 28 metric sets across all four subjects. The $\downarrow$ symbol 
indicates that the improvement is statistically significant. 
The significant improvement occurs to 
20 out of 28 metric sets across all four subjects
by amount ranging from 5.7\% to 25.7\%. 
The lowest level of improvement 
occurred on SPM for OpenOffice, while the highest levels of
occurred on \Name{} for MySQL.
Overall, these results  
suggest that {\em the use of 
feature selection improves the performance of the concurrency fault
prediction}.

\subsubsection{RQ1.2: Effectiveness of Static Metrics}
As Table~\ref{tab:metrics} shows, 
comparing the static concurrency metric set
(CStaMC) to the static sequential metric set
(SStaMC), when averaging the F-measures, CStaMC
improved over SStaMC ranging from 23.8\% to 64.2\%
on all four subjects. The best improvement is
for Apache and the worst improvement is for OpenOffice.
Table~\ref{tab:cc} (columns under the SStaMC) 
shows the improvements were 
statistically significant. 
The results indicate that \emph{when choosing static metrics
for predicting concurrency faults, CStaMC is better than SStaMC}.

When comparing CStaMC to \Name{}, 
\Name{} was more effective for all four subjects. 
Their F-measures are statistically significant. 
In other words, adding dynamic metrics to CStaMC improved the prediction
performance of static metrics. 
The effectiveness improvement achieved by
\Name{} with respect to CStaMC
ranged from 10.8\% (MySQL) to 16.2\%
(Apache). These results
indicate that \emph{the combination of
static and dynamic metric sets is more effective
than the static metric set alone}.
The results also implies that 
\emph{the use of the dynamic metric sets amplifies the
effectiveness of the defect prediction}.  

When comparing CStaMC to CDyMC, CDyMC performs
better than CStaMC, ranging from 2\% (Mozilla) to 7.2\% (OpenOffice).
However, such differences were statistically 
significant on only MySQL and OpenOffice. The
results imply that \emph{dynamic metrics have the potential
to improve the performance of prediction over
static metrics on certain programs}.

\subsubsection{RQ1.3: Effectiveness of Dynamic Metrics}
As shown in Table~\ref{tab:metrics} and 
Table~\ref{tab:cc} (columns under CDyMC), 
\Name{} was more effective than CDyMC on all 
four subjects. The improvements, ranging from 
3.9\% (MySQL) to 9.9\% (Apache), were  statistically 
significant. These results indicate that 
\emph{the combination of static metrics and
dynamic metrics has better performance than
the dynamic metrics alone,} and 
\emph{the use of the static metrics validates the prediction 
performance of \Name{}}. 

When comparing CDyMC and SDyMC, CDyMC could consistently improve
the performance of SDyMC on all subjects
ranging from  14.3\% (Mozilla) to 66.7\% (Apache).
The improvements were still statistically significant. 
These results imply that 
\emph{when choosing dynamic metrics
for defect prediction on concurrent programs, 
CDyMC is better than SStaMC}.

\subsubsection{RQ1.4: Different Machine Learners}
\label{subsub:rq1.3}
Table~\ref{tab:metrics} shows the performance of each set of metrics 
in different machine
learners in terms of the F-measure distributions from 100 times 
10-fold cross validations. 
As the figure shows, the F-measure distributions from different machine
learners did vary.  However, the trend indicates that \Name{}
performs better than the other metric sets over all learners.
Table~\ref{tab:ml} 
shows the comparisons among different learners across \Name{}
and the two metric sets used in  \Name{}. The numbers are
the effect sizes and those marked
as bold indicate that the differences are statistically 
significant different.

Among the mean values, Random Forest was significantly better 
than two other learners over all the metric sets.  
Only on \Name{} was Random Forest not significantly different
from Decision Tree. 
Therefore, we employed Random Forest in 
the other experiments mentioned in Section~\ref{subsec:vars}.
In contrast, Logistic
Regression was the worst choice for model construction in our experiment. 
%F-measure performance of
%distributions from Logistic Regression skewed a lot to zero. 
%Interestingly, Naive Bayesian was not the best choice in model 
%construction but the best one for SPM.

If these results generalize to other real subjects, 
we then conclude that {\em when constructing machine learning
models to predict concurrency faults, Random Forest is the best choice,
whereas Logistic Regression is inferior.}

\ignore{
\begin{figure*}[t!]
%\hspace{5pt}
\includegraphics[scale=0.5]{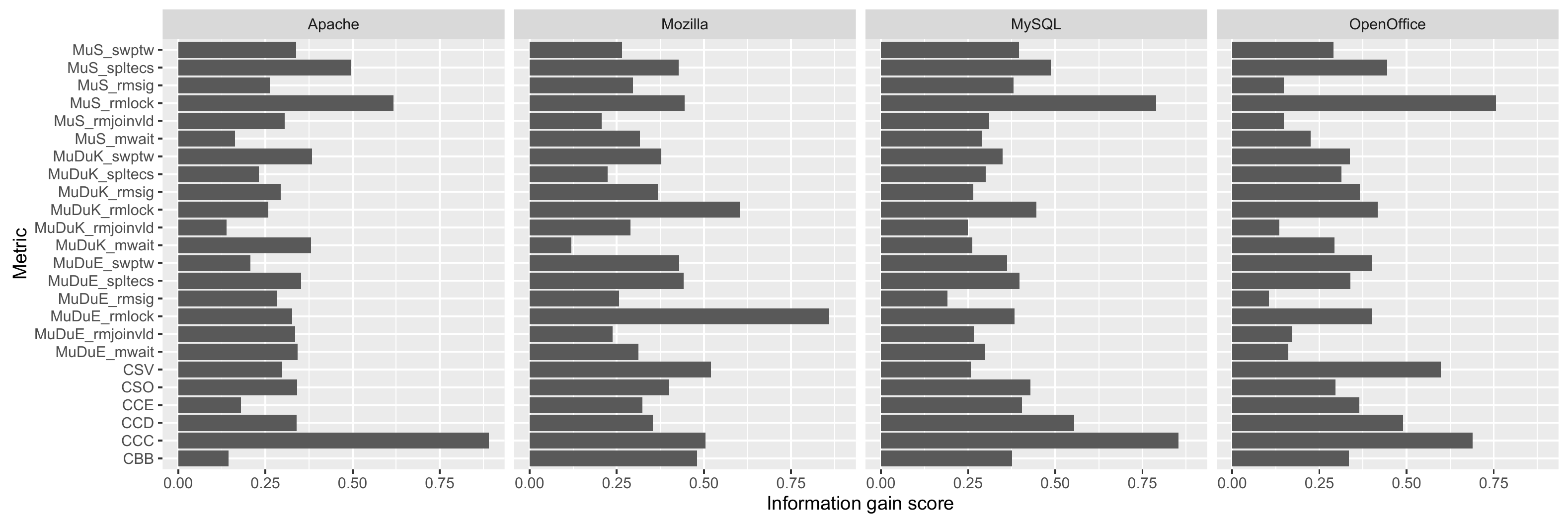}
\vspace*{-6pt}
\centering{
\caption{\textbf{\small \Name{} Metrics based on the Information Gain Ratio.}}
\label{fig:rank_individual}
}
%\vspace*{-15pt}
\end{figure*}

\begin{figure}[t!]
%\hspace{5pt}
\includegraphics[scale=0.51]{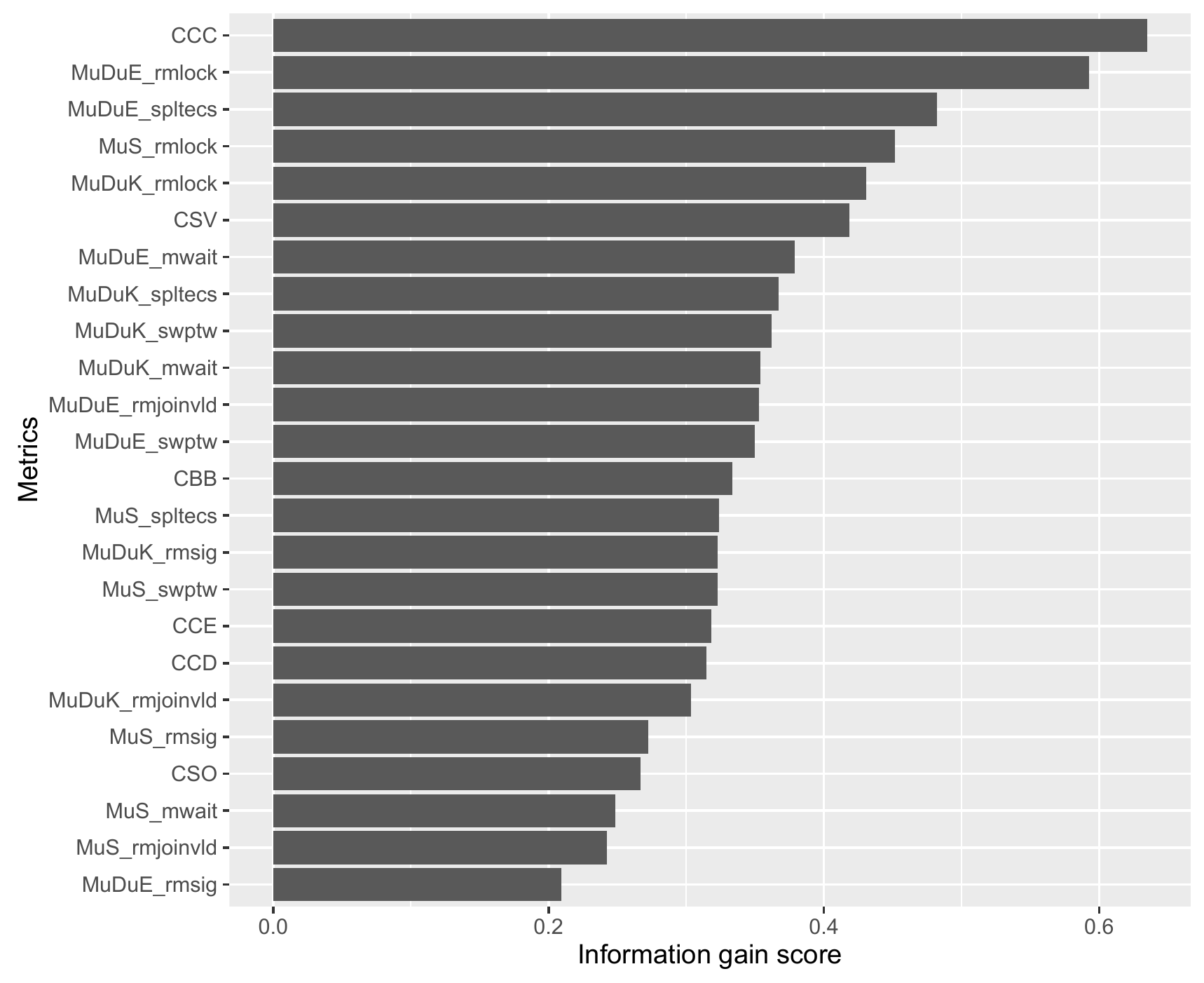}
\vspace*{-6pt}
\centering{
\caption{\textbf{\small \Name{} Metrics based on the Information Gain Ratio across
all subjects.}}
\label{fig:rank_all}
}
%\vspace*{-15pt}
\end{figure}

}

\subsection{RQ2: Metrics Effectiveness Analysis}

Table~\ref{tab:rank}  shows the Scott-Knott test results.  
The importance values of metrics in one group  
are statistically significantly different from those of metrics
in other groups. 
The results show that CCC (concurrency code complexity), MuDuE$_{spltecs}$, 
and MuDuS$_{rmlock}$ are the top three 
most important features that influence our random forest model. 
The best predictor was CCC. 
For code quality prediction, this metric resembles McCabe
complexity for sequential programs.  A previous study~\cite{Menzies10}
has shown that McCabe is effective at predicting defects
in sequential programs. Moreover, lock operations are 
commonly used  in concurrent programming to synchronize
shared resource accesses. These results
suggest subjects in our dataset that have higher 
concurrency code complexity, with more locks are more likely
to discriminate faulty instances from non-faulty instances. 

We next use Wilcoxon rank-sum and Cliff's delta to 
evaluate the effects of feature importance. 
Table~\ref{tab:effects} shows the factors that have p - value $<$ 0.01
and d $>$ 0.147 (i.e., statistically significant difference with at least a small effect size). 
We find that the faulty and non-faulty instances have statistically 
significant differences with at least a small effect size in 7 out of 
the 17 selected features. The effect sizes are small for 
most of the 17 factors, 
except for CCC, MuDuE$_{spltecs}$, 
and MuDuS$_{rmlock}$. The results are consistent with feature ranking
in Table~\ref{tab:rank}.

For the static metric set, three out of six metrics can differentiate faulty instances 
from non-faulty instances. Typically, when a program has more 
complex code, shared variables, and locks,  it becomes
more difficult to maintain the program and thus
is more likely to cause field failures. 
For the dynamic metric set, four out of seven metrics can 
differentiate faulty instances  from non-faulty instances. 
These metrics are all related to lock operations, which again suggest
that faulty instances are more likely to misuse locks.

\ignore{

The figure also shows that among the top 10
metrics for classification models, 
seven metrics are from the dynamic metric set.  
This suggests that {\em both static and dynamic metrics can play an important 
role in defect prediction,
and the quality of the dynamic metrics might be more important}.

Our results also suggest that different systems
require different metrics for best 
performance. For example MuDuE\_mlock is the most effective
metric in Mozilla, but not in the other subjects. 
By further examining the programs, we found that lock operations
are the major synchronizations used in Mozilla and also got
executed frequently.  
}

\subsection{RQ3: Across Projects Prediction}

RQ3 investigates whether a predictor for one application
group (dataset) can be used for other applications. 
We applied classification model built
by \Name{} from each dataset
to the instances of each of the other four  datasets. 
We then checked how accurate the prediction is by 
assessing the performance of each model. 
The prediction results are shown in  columns 2-5 of Table~\ref{tab:cp}. 
For example, columns 3-5 show the performance values
of using models learned from {\sc MySQL}, {\sc Mozzila}
and {\sc OpenOffice} to predict {\sc Apache}.

As the results show, 
the F-measure values for three out of 12 models were greater than 0.5, 
and six out of 12 models were greater than 0.4. 
%indicating that the cross project classification is effective.
%Only the BayesNet model learned from {\sc Small} had
%low effectiveness (F-measure = 0.359) in predicting {\sc mysql2}. 
Two out three models that are greater than 0.5 happen 
between {\sc Mozzila} and {\sc Apache}.
The two projects have certain similarities because they involve web
applications (i.e., in the case when Mozilla is used to predict Apache and in the
case when Apache is used to predict Mozilla).
%In addition, it is not surprising
%to see that the models are even more successful across similar 
%projects (i.e., {\sc mysql1} and {\sc mysql2}).

We next compare the across project prediction performance of \Name{}
to that of SPM. Columns 6-9 of Table~\ref{tab:cp} shows the F-measures
using SPM. The results indicate that \Name{} is more effective than
SPM on all 12 program pairs. 
The numbers marked as bold indicate that the differences 
are statistically significant. 

Overall, these results suggest that {\em \Name{} is moderately effective at
predicting concurrency faults across projects, and is more effective 
when predicting across similar projects. When comparing 
\Name{} to SPM, \Name{} is more effective at predicting concurrency 
faults across projects}. 
For future work, we intend to improve \Name{}
on cross-project defect prediction by applying 
Transfer Component Analysis+ (TCA+) analysis~\cite{nam2013transfer}.

\section{Summary and Discussion}
\label{discussion}

As presented in the previous section, we were able to 
demonstrate that \Name{} is useful for 
predicting defects of concurrent programs.  
Specifically, we showed (subject to stated 
threats to validity) that 
1) the metrics used in \Name{} are more effective at
predicting concurrency faults than sequential metrics,
2) the combination of static and dynamic metrics
has better performance than either the static metric set
or the dynamic metric set,
3) \Name{}'s dynamic concurrency metrics are
more effective contributors than static metrics 
based on the information ratios,
4) the concurrency code complexity metric (CCC) is the most
effective code metric, and
5) the models built on three 
datasets can be used to predict concurrency faults for the fourth dataset with 
good effectiveness.

One of our primary findings is that mutation metrics can significantly improve predictive performance and with large effect sizes. 
This is the first time any kind of concurrency mutation has been used to support fault prediction. Naturally, subsequent 
studies can further investigate/exploit this predictive improvement, perhaps in combination with other sets 
of metrics (e.g., process metrics). In this first study we present the empirical evidence that concurrency mutation 
metrics' effect size on prediction outcomes can be large, thereby motivating and opening up this avenue of research.

Our results have implications for practitioners and researchers, discussed
below.

\subsection{Implications for Practitioners}

Results indicate that concurrency-related
program metrics can be effective and are more effective than 
predictors learned from sequential program attributes.  
Practitioners can apply this finding by building a CCFG, obtaining the 
concurrency metrics (we plan to provide a tool in the future to simplify this), 
and substituting their metrics into our learned model in order to predict 
defect-prone functions.  In addition, our results showed that several static
code metrics (e.g., CCC, CSV) were good predictors.  
For example,  
industry practitioners can use the CCFG to calculate CCC and examine its distribution 
for their project.  Functions that have higher values of CCC should be examined and 
possibly subjected to additional code review and/or unit testing to lower the risk 
of concurrency faults.  Also, functions that share communication edges with
defect-prone functions warrant additional attention during software
assurance activities.

\subsection{Implications for Researchers}

\subB{The use of CCFG.}
Our study shows that code and mutation metrics can be used to predict faults
specific to
concurrent programs.  Researchers should consider adding the CCFG to their 
arsenal of program representations.  The \Name{} metrics may have other
applications, such as for predicting testability of concurrent
programs, predicting change prone components in concurrent programs, and predicting 
the number of tests 
that are needed to achieve coverage of concurrent programs.   

\subB{Lowering the recall.}
We inspected a few cases where an instance (i.e., function)
is mislabeled as non-faulty. This is because
a concurrency fault may involve more than one function because
the conflicting shared variables may exist in different
functions. A function labeled as non-faulty may
have conflicting accesses with other functions that are labeled
as faulty. To address this problem, we could employ
the following heuristic: if one function $f$ is labeled as faulty, 
other functions that can be reached through a communication
edge from $f$ should also be flagged as \emph{possible} faulty.
Identification and test of functions that can reach defect-prone functions 
may hold promise for improving the recall of concurrency
fault prediction and could be the focus 
of future work.

\subB{Incorporating test suite metrics.}
Our results suggest that dynamic metrics are more important
than static metrics in concurrency fault detection. 
This implies that the quality of the test suite is also a considerable 
factor in building effective predictors.  
In this work, the dynamic metric set
includes only mutation metrics. 
More dynamic metrics are worth being studied
to improve the performance of \Name{}. One direction
is to develop test suite metrics by which their executions
provide various observable attributes. 
For example, 
coverage metrics are commonly used as they 
directly measure the relationship
between the test suite and source code. 
Thus, predictors can be built by incorporating the 
coverage metrics of the
program under test.

Additionally, 
the finding on communication edges implies that researchers should examine 
concurrent edge coverage more carefully as an important test criterion. 
The {\em interleaving coverage criteria} has been widely
used to measure test suite quality for concurrent programs~\cite{Hong12, Bron05, Lu07}. 
An interleaving criterion is a pattern of inter-thread dependencies through $SV$ accesses
that helps select representative interleavings to effectively expose concurrency faults. 
An interleaving criterion is satisfied if all feasible 
interleavings of $SV$ defined in the criteria are covered. 
As part of the future work, we can employ a Def-Use criterion, which 
is satisfied if and only if a write {\tt w}
in one thread happens before a read {\tt r} in another thread
and there is no other write to the variable read by {\tt r} between them.
In fact, the Def-Use criterion is equivalent to communication edge coverage
in the context of CCFG.

\section{Related Work}
\label{related}

There has been much research on developing various software
metrics and prediction algorithms
to assess software quality~\cite{Fenton98, li2006experiences,Lessmann08,Lee11,Menzies10,Menzies07,
Basili96,Kim08,Moser08,WORDS-2005}.
For example, Lee et al.~\cite{Lee11} proposed 
a set of micro-interaction metrics (MIMs) that leverage developers' interaction
information combined with source code metrics to predict defects. 
Meneely~\cite{Meneely08} et al. examine structure of developer collaboration
and use developer network derived from code information to predict defects. 
Basili et al.~\cite{Basili96} used Chidamber and Kemerer metrics, and Ohlsson et al. ~\cite{ohlsson1996predicting} 
used McCabe's cyclomatic complexity for defect prediction.
Koru and Liu utilized static software measures along with defect data at
the class level to predict bugs using machine learning.
Menzies et al.~\cite{Menzies07} conclude that
static code metrics are useful in predicting defects under specific
learning methods. 
These techniques, however, focus on sequential
programs while ignoring code attributes and testability for
concurrent programs. 

Besides code metrics, other metrics can be obtained from different software artifacts.
For example, Ohlsson et al.~\cite{ohlsson1996predicting} study
metrics derived from design documents that are used to predict 
fault-prone modules. Metrics collecting from version control 
systems (e.g., number of commits) have also been used
to predict faults~\cite{herzig2013predicting,kim2007predicting,moser2008comparative}.
Nagappan et al.~\cite{nagappan2008influence} build fault 
prediction models by considering the organizational metrics. 
However, none of these techniques consider metrics or faults in concurrent
programs. In addition, they do not consider dynamic metrics
related to test executions.

%Machine learning techniques are popular for predicting software
%defect~\cite{Basili96,Kim08,Moser08,WORDS-2005}. 
%For example, Menzies et al.~\cite{Menzies07} conclude that
%static code metrics are useful in predicting defects under specific
%learning methods. 
%Khoshgoftaar et al.~\cite{Khoshgoftaar00}
%use a neural network to learn models from source code metrics
%to predict testability based on mutation analysis. 
%Jalbert et al.~\cite{Jalbert12}
%predict mutation scores by using source code metrics combined
%with coverage information. These techniques, however, focus on sequential
%programs while ignoring code attributes and testability for
%concurrent programs. 

The only related work found is the work by Mathews 
et al.~\cite{Mathews95} based on Ada programs. 
This work considers only the number of
synchronizations and conditional branches
that contain synchronization points without utilizing them 
to perform defect prediction.
However, this work does not aim to predict concurrency faults.
In addition, this work does not consider dynamic metrics.

%across-project defect predction
There have been many approaches to improving performance of
cross-project defect prediction~\cite{nam2013transfer, nam2017heterogeneous,ma2012transfer,
rahman2012recalling,turhan2012dataset,turhan2009relative, zhang2014towards}. 
For example,  Nam et al.~\cite{nam2013transfer}
adapted a state-of-the-art transfer learning technique called 
Transfer Component Analysis (TCA) and proposed TCA+. 
Turhan et al. proposed the nearest-neighbor (NN) filter to improve the 
performance of cross-project defect prediction~\cite{turhan2012dataset}. 
The basic idea of the NN filter is that prediction models are built by 
source instances that are nearest-neighbors of target instances.
In the future, we intend to leverage these techniques
to improve the performance of \Name{} in cross-project
concurrency fault prediction.

Various data quality issues can arise when constructing defect prediction datasets.
a variety of methods have been proposed for dealing with class imbalance problems.
Chawla et al.~\cite{chawla2002smote} proposed an over-sampling approach
in which the minority class is over-sampled by creating ``synthetic" instances 
rather than by over-sampling with replacement. 
Estabrooks et al.~\cite{estabrooks2004multiple}
and Barandela et al.~\cite{barandela2004imbalanced}
both suggested that a combination of over-sampling and under-sampling 
might be more effective to solve the class imbalance problem.
\Name{} employs an over-sampling technique~\cite{tan2015online}, i.e., SMOTE.

Studies have also shown that the presence of systematic data noise and bias in several open 
source data sets affect the performance of defect prediction models~\cite{bird2009fair, bachmann2010missing, nguyen2010case}. Therefore, research on improving mappings between bug reports and
code has been proposed. For example, Bird et al.~\cite{bird2010linkster}
develop a technique to manually annotate bug reports and code changes to 
reduce the overhead of manual data point inspection. Wu et al. ~\cite{wu2011relink}
propose an automatic link recovery algorithm 
to recover missing links between bug reports and code changes. 
Kim et al.~\cite{kim2011dealing} propose guidelines about
acceptable noise levels and propose a noise detection and elimination algorithm.
In the future, we will leverage the above techniques to speed
up the construction of datasets.

%Although prediction of defects of concurrent programs
%has not been previously researched, the related
%topic of using software metrics to predict faults in source code has been well researched.
%For example, Koru and Liu utilized static software measures along with defect data at
%the class level to predict bugs using machine learning.

There has been a great deal of research on mutation testing
for sequential programs~\cite{Ma06,Schuler09,Delamaro01}. 
Jia and  Harman~\cite{Jia11} provide a recent survey.
In this work, we focus on techniques that 
share similarities with ours. There has also been some work
on mutation testing for concurrent programs~\cite{Bradbury06, Gligoric13, Kaiser11}, 
which has been discussed in Section~\ref{background}. 
Other tools such as MuTMuT~\cite{Gligoric10} have been used
to optimize the execution of  mutants by reducing interleaving
space that has to be explored.  None of the techniques, however, attempt to 
predict defects  using software metrics. 

Recent work by Bowes et al.~\cite{Bowes2016} 
propose a mutation-aware fault prediction technique, 
which leverages guidance from mutation analysis 
to construct dynamic metrics. Both their work
and \Name{} use mutants and 
the test cases that cover and detect them for building
additional metrics. However, their technique focuses
on sequential programs. As shown in our results,
either static or dynamic sequential metrics 
are significantly less effective than 
\Name{}'s metrics.

%\vspace*{-10pt}
\section{conclusions}
\label{conclusions}

This paper presents an approach to predict defects
of concurrent programs at the function level. We proposed
six novel static code metrics, six static mutation
metrics, 12 dynamic mutation metrics, and combined them with a
dynamic test suite metric to learn four prediction models. 
We applied the models to four large-scale real-world programs. 
We found that \Name{} can outperform traditional defect 
prediction using sequential metrics. 
In all cases, both dynamic and static metrics feature prominently in the top 10
most influential metrics across all subjects,
providing consistent evidence that they are beneficial to 
the performance of prediction models.  
In addition, \Name{} showed good effectiveness when applied
to different software projects. 
Our study extended existing knowledge in the field of software quality 
metrics by proposing novel metrics specific to concurrent programs.
In the future, we will consider  other 
code metrics, test suite attributes, and  investigate their effectiveness as discussed in Section~\ref{discussion}.
We will  also study the effectiveness of  combining both sequential and concurrency
metrics for predicting both sequential and concurrency faults. 

\section*{Acknowledgments}

This work was supported in part by NSF grants CCF-1464032,  CCF-1652149, and CCF-1511117.

\bibliographystyle{abbrv}

\begin{flushleft}
\bibliography{bib/paper}
\end{flushleft}


\begin{IEEEbiography}[{\includegraphics[width=1in,height=1.25in,clip,keepaspectratio]{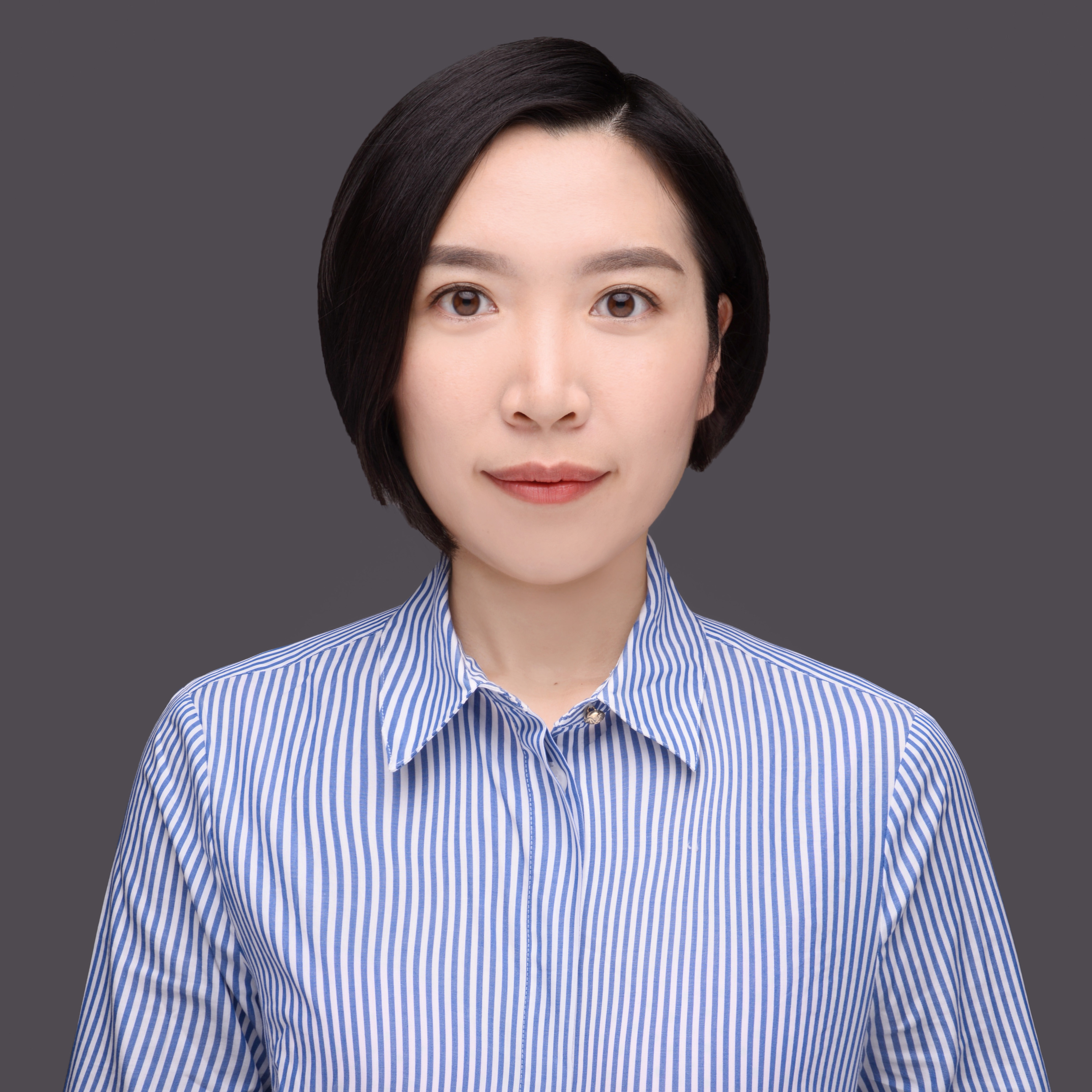}}]{Tingting Yu}
Tingting Yu is an Assistant Professor of Computer Science at University of Kentucky.  She received her M.S. and Ph.D degree from University of Nebraska-Lincoln in 2014, and B.E. degree in Software Engineering from Sichuan University in 2008. Her research is in software engineering, with focus on developing methods and tools for improving reliability and security of complex software systems; testing for sequential and concurrent software; regression testing; and performance testing. She is a member of the IEEE Computer Society.
\end{IEEEbiography}

\begin{IEEEbiography}[{\includegraphics[width=1in,height=1.25in,clip,keepaspectratio]{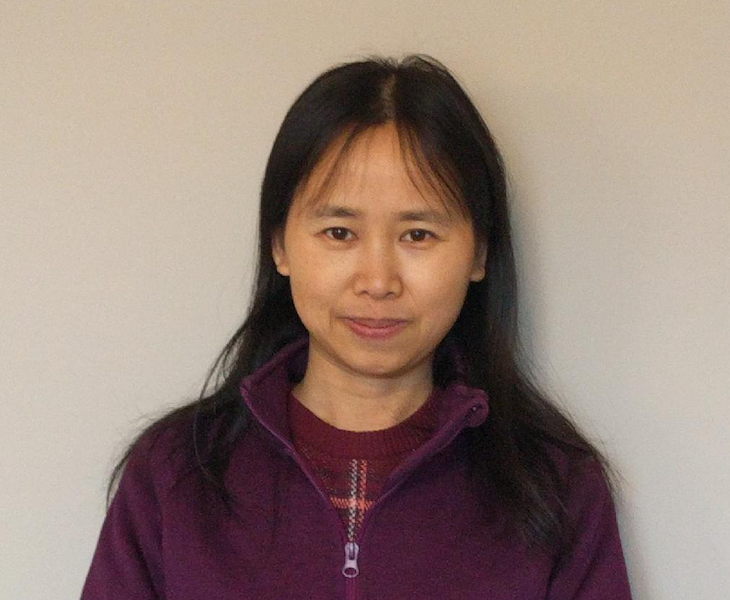}}]{Wei Wen}
Ms. Wei Wen received her M.S. degrees in Electrical \& Computer Engineering  and Computer Science from University of Kentucky in  2009 and 2016. She is currently working as a Mobile Software Engineer developing apps for both Android and iOS platforms. 
\end{IEEEbiography}

%\newpage

\begin{IEEEbiography}[{\includegraphics[width=1in,height=1.25in,clip,keepaspectratio]{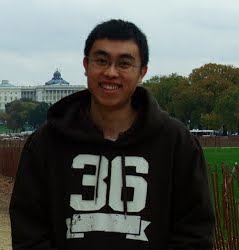}}]{Xue Han}
Mr. Xue Han is in his 4th year of Ph.D. study in the Computer Science Department, University of Kentucky. He is passionate about doing research in Software Testing, Search Based Software Testing, Performance Modeling, Program Analysis, Machine Learning, Natural Language Processing, and Data Mining. He has published several research papers in premier Software Engineering conferences. Before his Ph.D. study, he was working as a .NET Engineer. 
\end{IEEEbiography}

\begin{IEEEbiography}[{\includegraphics[width=1in,height=1.25in,clip,keepaspectratio]{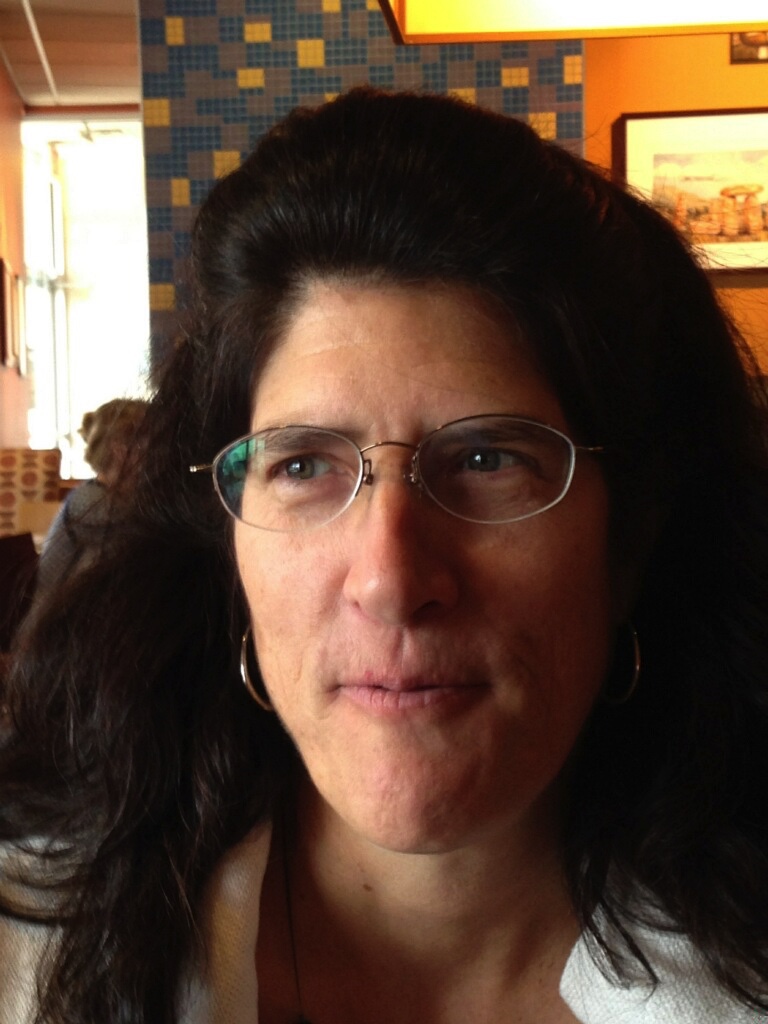}}]{Jane Huffman Hayes}
Jane Huffman Hayes received the PhD degree in information technology from George Mason University. She is a professor in the Department of Computer Science at the University of Kentucky. Her research interests include requirements, software verification and validation, traceability, maintainability, and reliability. She is a member of the IEEE Computer Society.
\end{IEEEbiography}

\end{document}